\pdfoutput=1
\documentclass[10pt,a4paper,twocolumn]{article}

\usepackage[a4paper,top=17.5mm,bottom=18.5mm,left=14.5mm,right=14.5mm,columnsep=6.2mm]{geometry}
\usepackage{mathpazo}
\usepackage[scaled=0.90]{helvet}
\usepackage{amsmath,amssymb,amsthm,mathtools}
\usepackage{graphicx}
\usepackage{booktabs,tabularx}
\usepackage[font=footnotesize,labelfont=bf]{caption}
\usepackage{xcolor}
\usepackage{microtype}
\usepackage{fancyhdr}
\usepackage{cite}
\usepackage{hyperref}

\newcommand{\PatentAppNo}{64/199,322}

\definecolor{primary}{RGB}{0,90,156}
\definecolor{slate}{RGB}{71,85,105}
\definecolor{calloutbg}{RGB}{240,247,255}

\hypersetup{
  colorlinks=true,
  linkcolor=primary,
  citecolor=primary,
  urlcolor=primary,
  pdftitle={Full-chip constrained quantum annealing via self-calibrating spin-susceptibility control},
  pdfauthor={Chia-Ho Ou}
}

\newtheorem{proposition}{Proposition}
\begin{document}

\twocolumn[{%
  \begin{@twocolumnfalse}
    \noindent{\footnotesize\textsf{\textbf{\textcolor{primary}{ARTICLE \enspace$\cdot$\enspace QUANTUM PHYSICS \& COMPUTATIONAL OPTIMIZATION}}}}\\[5pt]
    {\LARGE\bfseries Full-chip constrained quantum annealing via self-calibrating spin-susceptibility control\par}
    \vspace{7pt}
    {\normalsize\bfseries Chia-Ho Ou$^{1,2,*}$\par}
    \vspace{3.5pt}
    {\footnotesize\color{slate}
    $^1$Department of Computer Science and Information Engineering, National Pingtung University, Taiwan.\\
    $^2$Graduate School of Information Sciences, Tohoku University, Japan.\\
    $^*$e-mail: \href{mailto:cho@mail.nptu.edu.tw}{cho@mail.nptu.edu.tw}\par}
    \vspace{7pt}
    \noindent\colorbox{calloutbg}{%
      \parbox{\dimexpr\textwidth-2\fboxsep\relax}{%
        \vspace{4.5pt}
        \noindent{\small\textbf{\color{primary}Abstract}}\\[2.5pt]
        {\small Constrained binary optimization on sparsely connected superconducting quantum annealers is limited by quadratic penalties, which induce dense couplings requiring minor embedding, and scalar dual updates stall under overlapping constraints. This study presents a closed-loop primal--dual scheme, designated as Self-Calibrating Quantum Susceptibility Quasi-Newton (\texttt{SelfCal-QSQN}), that encodes linear equalities and inequalities in longitudinal fields without penalty couplers or slack qubits. Each iteration estimates the cross-constraint susceptibility from the covariance of a single batch of reads and updates its gain online without perturbation queries. In each of four full-chip tests (four or eight constraints on the D-Wave Zephyr and Pegasus processors with $4{,}560$ and $5{,}612$ spins, respectively), \texttt{SelfCal-QSQN} yields a lower final ensemble residual than both calibrated first-order baselines on all six seed-matched instances. In the best case, Zephyr with four constraints, the mean residuals are $0.092$ versus $0.885$ and $1.015$. In that case, a greedy 1-bit polish with the best state over all iterations retained reaches exact feasibility in $5/6$ runs, compared with $3/6$ and $2/6$ for the baselines.}
        \vspace{4.5pt}
      }%
    }
    \vspace{10pt}
  \end{@twocolumnfalse}
}]

\section{Introduction}

Recent advancements in physical Ising machines and superconducting flux-qubit processors have facilitated the development of hardware-accelerated discrete optimization, in which low-energy spin configurations are sampled across rugged energy landscapes \cite{kadowaki1998qa,farhi2001adiabatic,santoro2002martonak,johnson2011qa,albash2018aqc,hauke2020perspectives,mohseni2022ising,inagaki2016cim,goto2019sb,aramon2019da}. Practical decision tasks arise in a wide variety of real-world applications, including quantitative portfolio allocation, electric power-grid commitment, logistics routing, molecular fragment selection, and so on. Such tasks rarely reduce to unconstrained spin glasses since the candidate binary configurations are required to satisfy coupled resource budgets, cardinality targets, and capacity inequalities simultaneously \cite{lucas2014ising}. Successive generations of superconducting quantum annealing processors have expanded the processor scale and the coupler degree from Pegasus ($d=15$, \texttt{Advantage}) to Zephyr ($d=20$, \texttt{Advantage2}) \cite{boothby2020pegasus,dwave_zephyr2021,dwave_adv_perf,dwave_adv2_4400q,king2023coherent,king2025beyond}. However, the solution of constrained binary programs on sparse hardware lattices is limited by the cost of embedding dense penalty couplings \cite{choi2008minor,bian2014sparse,venturelli2015sk,pmc_clique2020,dwave_embedding_doc}.

\begin{figure*}[!htbp]
  \centering
  \includegraphics[width=\textwidth]{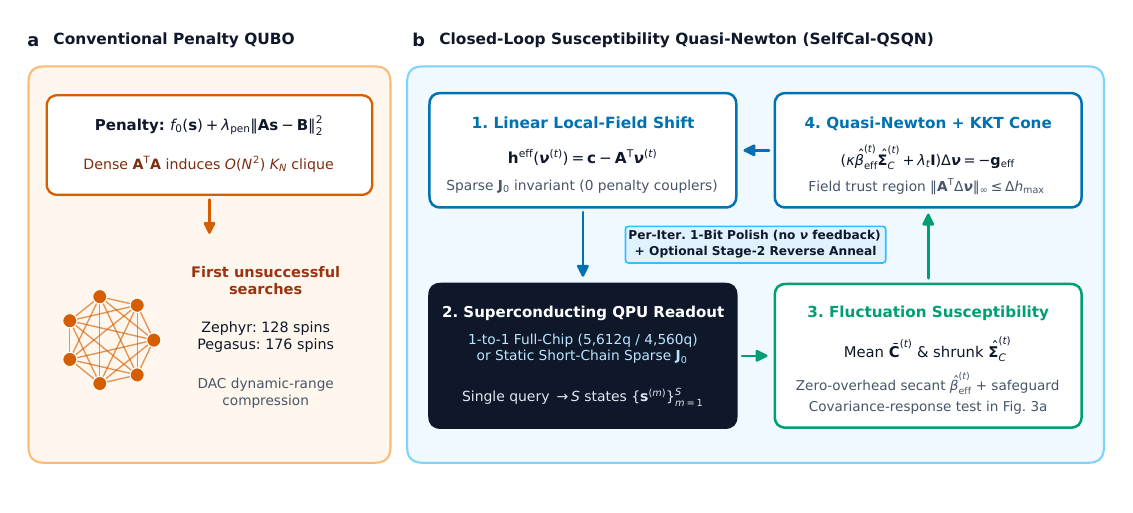}
  \caption{\textbf{Decoupling of algebraic constraints from the physical coupler topology via \texttt{SelfCal-QSQN}.} \textbf{a}, In the conventional quadratic-penalty QUBO formulation, the squared constraints generate dense $O(N^2)$ all-to-all $\mathbf{A}^{\mathsf{T}}\mathbf{A}$ cliques that require minor embedding and compress the analog DAC dynamic range. \textbf{b}, In the closed-loop \texttt{SelfCal-QSQN} scheme, the linear equalities and active-set inequalities are absorbed strictly into the longitudinal fields $\mathbf{h}^{\mathrm{eff}}(\boldsymbol{\nu}^{(t)}) = \mathbf{c} - \mathbf{A}^{\mathsf{T}}\boldsymbol{\nu}^{(t)}$ such that the sparse $\mathbf{J}_0$ couplers remain invariant, and the dual curvature is reconstructed from the single-shot QPU fluctuation covariances $\hat{\boldsymbol{\Sigma}}_C^{(t)}$ and the online secant freeze-out self-calibration $\hat{\beta}_{\mathrm{eff}}^{(t)}$, followed by per-iteration 1-bit polish and optional Stage-2 reverse annealing. The label ``single query'' in \textbf{b} denotes one batch of $S$ reads under one programmed Hamiltonian per iteration.}
  \label{fig:concept}
\end{figure*}

Existing constraint-handling schemes for physical Ising machines and quantum annealers fall into three broad categories, namely static quadratic-penalty formulations, constraint-preserving driver schemes, and dual multiplier schemes. In quadratic unconstrained binary optimization (QUBO) penalty schemes \cite{lucas2014ising}, $K$ linear constraints $\mathbf{C}(\mathbf{s}) = \mathbf{A}\mathbf{s} - \mathbf{B} = \mathbf{0}$ over the spin variables $\mathbf{s} \in \{-1,+1\}^N$ are appended to the intrinsic objective $f_0(\mathbf{s}) = \frac{1}{2}\mathbf{s}^{\mathsf{T}}\mathbf{J}_0\mathbf{s} + \mathbf{c}^{\mathsf{T}}\mathbf{s}$ by means of a squared penalty $\lambda_{\mathrm{pen}}\|\mathbf{A}\mathbf{s} - \mathbf{B}\|_2^2$. Even if the intrinsic coupling graph $\mathbf{J}_0$ is sparse or hardware-native, the expansion $(\mathbf{A}\mathbf{s})^{\mathsf{T}}(\mathbf{A}\mathbf{s}) = \mathbf{s}^{\mathsf{T}}(\mathbf{A}^{\mathsf{T}}\mathbf{A})\mathbf{s}$ superimposes dense $O(N^2)$ pairwise couplings across all of the constrained spins (Fig.~\ref{fig:concept}a). The minor-embedding of the resulting complete graph $K_N$ onto bounded-degree Chimera and Pegasus lattices requires ferromagnetic qubit chains that consume $O(N^2)$ physical qubits \cite{pmc_clique2020,dwave_embedding_doc}. Although an ideal defect-free Zephyr $Z_{12}$ lattice theoretically supports cliques up to $K_{184}$ \cite{dwave_zephyr2021,zephyr_noise2025}, inoperable qubits and couplers on physical working graphs restrict practical embeddability. In the native benchmark, the tested procedure found no clique embedding at 176 or 256 spins on Pegasus or at 128, 176 or 256 spins on Zephyr. Larger sizes were skipped (see Methods). Even when an embedding is found, the penalty weight affects calibration. Specifically, large penalty weights $\lambda_{\mathrm{pen}}$ inflate ferromagnetic chain-strength requirements and compress $\mathbf{J}_0$ beneath the precision floor of the digital-to-analog converter (DAC) \cite{zephyr_noise2025,dwave_solver_doc}, whereas moderate penalty weights preserve $\mathbf{J}_0$ resolution at the cost of substantial constraint violations. In addition, long ferromagnetic chains exponentially suppress logical-subspace sampling weights and introduce post-processing projection biases into the readout statistics \cite{marshall2020perils}. Furthermore, binary slack encodings for the inequalities $\mathbf{A}_{\mathrm{ineq}}\mathbf{s} \le \mathbf{B}_{\mathrm{ineq}}$ increase the qubit overhead \cite{yonaga2020ineq}. Recent slack-free penalty constructions such as topology-aware Walsh--Fourier penalization (\texttt{TAWP}) \cite{lee2026tawp} project non-quadratic inequality surrogates directly onto hardware-native couplers without auxiliary slack qubits. However, such constructions remain static single-Hamiltonian penalty approximations that modify the two-body couplers rather than iteratively adjusting the multipliers. Constraint-preserving driver Hamiltonians and alternating-operator mixers confine the evolution to feasible subspaces without penalty terms \cite{hen2016cqa,hadfield2019qaoa}. However, the implementation of multi-body feasible-subspace drivers for general coupled linear equalities and inequalities requires non-native higher-order interactions, which are not available on standard transverse-field hardware. Conversely, Lagrangian dualization, cutting-plane outer approximation, and Hubbard--Stratonovich approaches \cite{ronagh2016lagrangian,yonaga2020ineq,ohzeki2020breaking,gabbassov2025lagrangian} relax the coupled constraints via dual multipliers $\boldsymbol{\nu} \in \mathbb{R}^K$, and absorb the linear constraints into the longitudinal local fields $\mathbf{h}^{\mathrm{eff}}(\boldsymbol{\nu}) = \mathbf{c} - \mathbf{A}^{\mathsf{T}}\boldsymbol{\nu}$ via first-order subgradient or outer-approximation steps. Unbalanced-penalization Ohzeki extensions (\texttt{UPOM} and \texttt{DU-UPOM}) \cite{hagiwara2026upom} likewise remove slack variables and squared-residual couplers for inequality constraints, and derive multi-constraint moment-covariance derivatives to train layer-wise scalar step-size schedules via deep unfolding. However, the numerical validation of these extensions has been restricted to classical Monte Carlo sampling on single-constraint knapsack instances without QPU experiments. In addition, related multiplier-based schemes have been developed for gate-model variational solvers, namely the multiblock alternating direction method of multipliers \cite{gambella2020admm} and the primal--dual variational quantum eigensolver with constraints \cite{le2024vqec}. Furthermore, under overlapping constraints ($K \ge 2$), scalar step sizes ignore the off-diagonal cross-constraint susceptibility, and thus face an acute conditioning dilemma. In other words, supercritical scalar steps diverge into DAC rail saturation, whereas calibrated fixed or diminishing scalar steps stall along ill-conditioned differential constraint directions.

The resolution of cross-constraint coupling requires second-order curvature information. However, textbook Boltzmann-machine Newton updates cannot be applied directly to physical quantum annealers as a result of three hardware non-idealities. First, the superconducting spin readout near the non-equilibrium freeze-out boundary \cite{amin2015searching,benedetti2016temp,raymond2016warming,marshall2017freezeout,marshall2019power,nelson2022gibbs} samples an instance- and field-dependent effective inverse temperature $\beta_{\mathrm{eff}}$ that is unknown a priori and shifts substantially across processor topologies and annealing durations, whereas offline multi-ensemble temperature estimators require auxiliary sampling overhead. Second, under practical read budgets $S$ comparable to $K$, the raw empirical covariance matrices suffer severe Marchenko--Pastur and Wishart small-eigenvalue collapse \cite{ledoit2004well}, and thus the quantum shot noise is amplified into unstable multiplier jumps. Third, unguarded dual updates induce abrupt per-qubit longitudinal field shifts that destabilize the spin-glass freeze-out landscape. Therefore, the accurate reconstruction of a well-conditioned dual susceptibility curvature on noisy analog annealers without additional circuit overhead remains an open problem.

Accordingly, this study proposes a hardware-native quantum--classical closed-loop primal--dual workflow that resolves all three of these bottlenecks with zero auxiliary quadratic penalty couplings while preserving the sparse coupler topology (Fig.~\ref{fig:concept}b). In the proposed workflow, designated as Self-Calibrating Quantum Susceptibility Quasi-Newton (\texttt{SelfCal-QSQN}), each outer iteration requires only a single batch of $S$ reads from the quantum processing unit (QPU) under one programmed Hamiltonian. This batch yields both the ensemble constraint residual $\bar{\mathbf{C}}^{(t)}$ and its shrinkage-regularized fluctuation covariance $\hat{\boldsymbol{\Sigma}}_C^{(t)}$ without any auxiliary perturbation queries. An effective susceptibility gain $\hat{\beta}_{\mathrm{eff}}^{(t)}$ (absorbing the unknown hardware freeze-out inverse temperature $\beta_{\mathrm{eff}}$ and covariance shrinkage scaling) is self-calibrated online from consecutive dual displacements by means of a signal-to-noise ratio (SNR)-gated secant estimator. In addition, closed-loop stability within the analog DAC bounds is enforced by means of residual-adaptive Levenberg--Marquardt damping, a non-monotone overshoot safeguard, a slack-free semi-smooth Karush--Kuhn--Tucker (KKT) cone projection, and a direction-preserving local-field trust region. The lowest-violation raw samples of each iteration are then passed through a lightweight deterministic 1-bit greedy polish ($O(NK)$ per flip), which is not fed back to the dual update. The performance of \texttt{SelfCal-QSQN} is evaluated by means of a series of $891$ multi-iteration hardware runs on the D-Wave \texttt{Advantage\_\allowbreak system6} and \texttt{Advantage2\_\allowbreak system1} processors. The hardware runs involve $6{,}850.36\text{ s}$ of pure QPU access time and $44{,}066$ Solver API (SAPI) quantum machine instructions (QMIs), as summarized in Table~\ref{tab:campaigns}. In the experiments, the performance of \texttt{SelfCal-QSQN} is compared to that of two existing quantum schemes, namely the quadratic-penalty QUBO scheme \cite{lucas2014ising} and the first-order dual scheme of Ohzeki \cite{ohzeki2020breaking}, as well as classical mixed-integer linear programming (MILP) and linear-programming (LP) baselines equipped with the identical 1-bit polish. Rather than claiming standalone asymptotic quantum speedup over classical optimization, the evaluation focuses on native topology preservation and hard-constraint feasibility convergence across full-chip lattices of up to $5{,}612$ physical qubits.

\begin{table}[!htbp]
  \centering
  \caption{\textbf{Summary of the $891$-run superconducting QPU experimental campaign.} Pure hardware QPU access times (\texttt{qpu\_access\_time}) and unique Solver API quantum machine instructions (SAPI QMIs) recorded on \texttt{Advantage\_system6} (Pegasus P16, up to $5{,}612\text{q}$) and \texttt{Advantage2\_system1} (Zephyr Z12, up to $4{,}560\text{q}$).}
  \label{tab:campaigns}
  \resizebox{\linewidth}{!}{%
    \begin{tabular}{@{}lrrr@{}}
      \toprule
      \textbf{Hardware Campaign Suite} & \textbf{QPU Runs} & \textbf{Unique QMIs} & \textbf{Pure QPU Time (s)} \\
      \midrule
      Phase 1: Native Scale-Up, Calibration, FDT \& Physics & 316 & 12,076 & 1,723.44 \\
      Phase 2: FDT, High-$K$, Schedule, Wishart, Full-Chip, Portfolio \& Zephyr Scale-Up & 213 & 14,286 & 2,207.76 \\
      Phase 3: Non-Native Graphs, Mixed KKT \& Zephyr $S$ & 94 & 3,940 & 536.59 \\
      Phase 4: Calibrated Ohzeki \& $\mathrm{SRT}=0$ response probes & 268 & 13,764 & 2,382.57 \\
      \midrule
      \textbf{Grand Total} & \textbf{891} & \textbf{44,066} & \textbf{6,850.36 (114.17 min)} \\
      \bottomrule
    \end{tabular}%
  }
\end{table}

\section{Results}

\subsection{Problem formulation and proposed scheme}

Consider a constrained discrete Ising optimization problem comprising $N$ binary spins $\mathbf{s} = [s_1, s_2, \dots, s_N]^{\mathsf{T}} \in \{-1,+1\}^N$ that interact over a sparse undirected edge set $\mathcal{E}$. Note that the problem can be mapped equivalently from the binary variables $\mathbf{x} = (\mathbf{s}+\mathbf{1})/2 \in \{0,1\}^N$ without altering the interaction sparsity. Let the symmetric zero-diagonal two-body coupling matrix supported on $\mathcal{E}$ be denoted by $\mathbf{J}_0 \in \mathbb{R}^{N \times N}$ and the intrinsic longitudinal bias vector be denoted by $\mathbf{c} \in \mathbb{R}^N$. It is assumed that the spin configuration is subject to $K_{\mathrm{eq}}$ linear equality constraints and $K_{\mathrm{ineq}}$ linear inequality constraints, where $K = K_{\mathrm{eq}} + K_{\mathrm{ineq}}$. To minimize the intrinsic Ising energy under the coupled resource constraints, the primal discrete optimization problem is formulated as
\begin{subequations}\label{eq:p1}
\begin{align}
  \text{(P1)} \quad \min_{\mathbf{s} \in \{-1,+1\}^N} \quad & f_0(\mathbf{s}) = \frac{1}{2}\mathbf{s}^{\mathsf{T}}\mathbf{J}_0\mathbf{s} + \mathbf{c}^{\mathsf{T}}\mathbf{s} \label{eq:p1_obj}\\
  \text{subject to} \quad & \mathbf{A}_{\mathrm{eq}}\mathbf{s} = \mathbf{B}_{\mathrm{eq}}, \label{eq:p1_eq}\\
  & \mathbf{A}_{\mathrm{ineq}}\mathbf{s} \le \mathbf{B}_{\mathrm{ineq}}, \label{eq:p1_ineq}
\end{align}
\end{subequations}
where $\mathbf{A}_{\mathrm{eq}} \in \mathbb{R}^{K_{\mathrm{eq}} \times N}$ and $\mathbf{B}_{\mathrm{eq}} \in \mathbb{R}^{K_{\mathrm{eq}}}$ specify the equality resource targets, and $\mathbf{A}_{\mathrm{ineq}} \in \mathbb{R}^{K_{\mathrm{ineq}} \times N}$ and $\mathbf{B}_{\mathrm{ineq}} \in \mathbb{R}^{K_{\mathrm{ineq}}}$ specify the upper capacity bounds. The objective of Problem~(P1) in equation~\eqref{eq:p1_obj} is to minimize the pairwise spin-interaction frustration and the local-field cost. Constraint~\eqref{eq:p1_eq} states that $K_{\mathrm{eq}}$ global or modular conservation laws hold with zero residual, and Constraint~\eqref{eq:p1_ineq} states that $K_{\mathrm{ineq}}$ sector or capacity budgets are not exceeded. Problem~(P1) belongs to the class of NP-hard non-convex binary quadratic programs. Furthermore, if constraints~\eqref{eq:p1_eq}--\eqref{eq:p1_ineq} are squared into QUBO penalties, the sparsity of $\mathcal{E}$ is destroyed by the resulting dense $\mathbf{A}^{\mathsf{T}}\mathbf{A}$ couplings.

To preserve the sparse graph $\mathcal{E}$, the constraints are stacked as $\mathbf{A} = [\mathbf{A}_{\mathrm{eq}}; \mathbf{A}_{\mathrm{ineq}}] \in \mathbb{R}^{K \times N}$ and $\mathbf{B} = [\mathbf{B}_{\mathrm{eq}}; \mathbf{B}_{\mathrm{ineq}}] \in \mathbb{R}^K$, the constraint residual vector is defined as $\mathbf{C}(\mathbf{s}) = \mathbf{A}\mathbf{s} - \mathbf{B}$, and a dual multiplier vector $\boldsymbol{\nu} \in \mathcal{K}^* = \mathbb{R}^{K_{\mathrm{eq}}} \times \mathbb{R}_{\le 0}^{K_{\mathrm{ineq}}}$ is introduced. Under an effective non-equilibrium freeze-out inverse temperature $\beta_{\mathrm{eff}} > 0$ \cite{amin2015searching}, Problem~(P1) is relaxed into the smooth concave dual free-energy maximization problem given by
\begin{equation}
  \begin{aligned}
    \text{(P2)} \quad \max_{\boldsymbol{\nu} \in \mathcal{K}^*} \quad &\Phi(\boldsymbol{\nu}) \\[-2pt]
    &= -\beta_{\mathrm{eff}}^{-1}\ln \sum_{\mathbf{s} \in \{-1,+1\}^N} \exp\bigl[-\beta_{\mathrm{eff}}H_{\mathrm{eff}}(\mathbf{s};\boldsymbol{\nu})\bigr],
  \end{aligned}
  \label{eq:p2}
\end{equation}
where $H_{\mathrm{eff}}(\mathbf{s};\boldsymbol{\nu}) = \frac{1}{2}\mathbf{s}^{\mathsf{T}}\mathbf{J}_0\mathbf{s} + \mathbf{h}^{\mathrm{eff}}(\boldsymbol{\nu})^{\mathsf{T}}\mathbf{s} + \boldsymbol{\nu}^{\mathsf{T}}\mathbf{B}$ denotes the dual-shifted Ising Hamiltonian, and $\mathbf{h}^{\mathrm{eff}}(\boldsymbol{\nu}) = \mathbf{c} - \mathbf{A}^{\mathsf{T}}\boldsymbol{\nu}$ denotes the effective longitudinal field vector. Note that $\mathbf{J}_0$ in Problem~(P2) remains strictly independent of $\boldsymbol{\nu}$ over all of the outer iterations. The exact dual curvature and the computational complexity of solving Problem~(P2) from single-shot spin ensembles are characterized by the following proposition. The proof is given in Methods.

\begin{proposition}[Fluctuation--Dissipation Susceptibility Curvature and Dual-Space Equivalence]\label{prop:fdt}
Under the freeze-out Gibbs measure $P_{\beta_{\mathrm{eff}}}(\mathbf{s};\boldsymbol{\nu}) \propto \exp[-\beta_{\mathrm{eff}}H_{\mathrm{eff}}(\mathbf{s};\boldsymbol{\nu})]$, the gradient and Hessian of $\Phi(\boldsymbol{\nu})$ in Problem~(P2) satisfy $\nabla_{\boldsymbol{\nu}}\Phi(\boldsymbol{\nu}) = -\bar{\mathbf{C}}(\boldsymbol{\nu})$ and $\nabla_{\boldsymbol{\nu}}^2\Phi(\boldsymbol{\nu}) = -\nabla_{\boldsymbol{\nu}}\bar{\mathbf{C}}(\boldsymbol{\nu}) = -\beta_{\mathrm{eff}}\boldsymbol{\Sigma}_C(\boldsymbol{\nu}) \preceq \mathbf{0}$, where $\bar{\mathbf{C}}(\boldsymbol{\nu}) = \langle\mathbf{C}(\mathbf{s})\rangle_{\boldsymbol{\nu}}$ and $\boldsymbol{\Sigma}_C(\boldsymbol{\nu}) = \mathrm{Cov}_{\boldsymbol{\nu}}[\mathbf{C}(\mathbf{s})]$. Furthermore, for any shrinkage weight $\alpha \in [0,1]$, positive gain $\kappa\hat{\beta}_{\mathrm{eff}}^{(t)} > 0$, positive damping $\lambda_t > 0$, and $S$-sample centered residual matrix $\tilde{\mathbf{C}} \in \mathbb{R}^{S \times K}$ with $S \ge 2$, the regularized quasi-Newton system $\mathbf{M}_t\Delta\boldsymbol{\nu}_{\mathrm{raw}}^{(t)} = -\mathbf{g}_{\mathrm{eff}}^{(t)}$ (with the curvature operator $\mathbf{M}_t$ and the active-set residual $\mathbf{g}_{\mathrm{eff}}^{(t)}$ defined in Methods) is strictly positive definite, yields a raw step $\Delta\boldsymbol{\nu}_{\mathrm{raw}}^{(t)}$ that is an ascent direction with respect to the sampled gradient estimate on the active KKT subspace whenever $\mathbf{g}_{\mathrm{eff}}^{(t)} \ne \mathbf{0}$, and, for the single-iteration shrunk covariance estimate (memory weight $\omega = 1$ in Methods), admits an exact $O(S^2 K + S^3)$ Sherman--Morrison--Woodbury dual solution that is identical to the solution obtained by $O(K^3)$ direct inversion.
\end{proposition}

Based on Proposition~\ref{prop:fdt}, \texttt{SelfCal-QSQN} performs a closed-loop update at each outer iteration $t$ (see Methods and Algorithm~1). First, $S$ spin reads are acquired under $\mathbf{h}^{\mathrm{eff}}(\boldsymbol{\nu}^{(t)})$. Second, the sample covariance is regularized by means of diagonal-target shrinkage to form $\hat{\boldsymbol{\Sigma}}_C^{(t)}$, with exponential moving average (EMA) smoothing applied only under low read budgets. Third, $\hat{\beta}_{\mathrm{eff}}^{(t)}$ is updated from the consecutive secant displacements $(\Delta\boldsymbol{\nu}^{(t-1)}, \Delta\bar{\mathbf{C}}^{(t)})$ with non-monotone overshoot safeguarding and zero additional QPU calls. Finally, $\Delta\boldsymbol{\nu}_{\mathrm{raw}}^{(t)}$ is contracted uniformly via $\tau_t = \min(1, \Delta h_{\max}/\|\mathbf{A}^{\mathsf{T}}\Delta\boldsymbol{\nu}_{\mathrm{raw}}^{(t)}\|_\infty)$ and is then projected onto $\mathcal{K}^*$.

\subsection{Full-chip native scaling and comparison with MILP}

\begin{figure*}[!htbp]
  \centering
  \includegraphics[width=\textwidth]{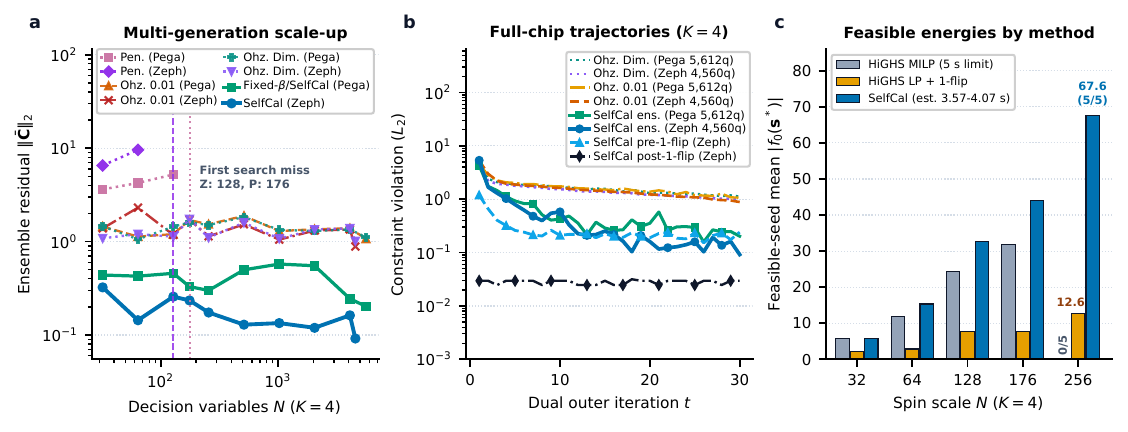}
  \caption{\textbf{Full-chip native scaling up to $5{,}612$ qubits, pre-polish raw QPU convergence, and classical comparisons under the stated budgets.} \textbf{a}, Raw QPU ensemble residual $\|\bar{\mathbf{C}}\|_2$ ($K=4$, mean across $5$ seeds per scale and $6$ seeds at full chip, with standard deviations listed in Table~\ref{tab:scaling}) for $N=32$ to $5{,}612$ on Pegasus and $N=32$ to $4{,}560$ on Zephyr. The first tested size with an unsuccessful clique search was 128 spins on Zephyr and 176 on Pegasus, and calibrated first-order Ohzeki baselines (measured on separately extracted instances for $N \le 4{,}096$, see Methods) stall near $\|\bar{\mathbf{C}}\|_2 \approx 0.9\text{--}2.3$. \textbf{b}, Full-chip ($K=4$) trajectories over $30$ outer iterations, plotting raw QPU ensemble residuals alongside the Zephyr \texttt{SelfCal-QSQN} pre-polish raw QPU minimum sample violation ($\min_m\|\mathbf{C}(\mathbf{s}_{\mathrm{raw}}^{(m,t)})\|_2$, with a per-seed trajectory minimum of $0.108 \pm 0.017$ across $6$ seeds at $N=4{,}560$) and its per-iteration post-polish violation (labeled 1-flip in the legend, whereas the $83.3\%$ exact-zero rate refers to the best incumbent across the $30$ polish passes). \textbf{c}, Comparison of \texttt{HiGHS} MILP ($5.0\text{-s}$ solver time limit on the reference host), \texttt{HiGHS} LP relaxation rounded and passed through the identical 1-bit polish over $22$ batches of $1{,}000$ draws (\texttt{HiGHS LP + 1-flip}), and \texttt{SelfCal-QSQN} on Zephyr ($3.57\text{--}4.07\text{ s}$ estimated compute-only time, including $3.56\text{--}4.04\text{ s}$ pure QPU access, outer-loop updates, and 1-bit polish) for $N=32\text{--}256$.}
  \label{fig:scaling}
\end{figure*}

The multi-generation native hardware scaling of the proposed workflow was first evaluated by means of a series of experiments on frustrated spin-glass objectives subject to $K=4$ overlapping equality constraints. The problem size was increased from $N=32$ to $100\%$ of the connected chip capacity, namely $N=5{,}612$ spins ($40{,}088$ couplers) on \texttt{Advantage\_\allowbreak system6} and $N=4{,}560$ spins ($41{,}205$ couplers) on \texttt{Advantage2\_\allowbreak system1}. Within each processor architecture, the same instance generator, problem sizes, seeds, read budgets, and outer iteration limits were applied to all of the compared iterative dual schemes, using $5$ independent seeds for $N \le 4{,}096$ ($S=1{,}000$ reads) and $6$ seeds at full chip ($S=2{,}000$ reads over $30$ iterations). Note that the calibrated \texttt{Ohzeki} baselines were executed on the identical problem instances at full chip, whereas for $N \le 4{,}096$ they were executed on separately extracted native subgraphs of the same sizes (see Methods). Therefore, the sub-full-chip comparisons with these two baselines are unpaired comparisons of multi-seed statistics. On Pegasus, the Phase-1 sub-full-chip sweep ($N \le 4{,}096$) executed $18$ iterations of \texttt{Fixed-}$\beta$\texttt{ QSQN} (defined in Methods), whereas full-chip Pegasus ($N=5{,}612$) and all ten Zephyr scales ($N=32\text{--}4{,}560$, $22$ iterations for $N \le 4{,}096$) executed \texttt{SelfCal-QSQN}. As shown in Fig.~\ref{fig:scaling}a and Table~\ref{tab:scaling}, the quadratic-penalty QUBO scheme ($\lambda_{\mathrm{pen}}=0.25$, selected via the hardware grid sweep in Methods) consumes $154$, $436$, and $1{,}640$ physical qubits at $N=32, 64,$ and $128$ on Pegasus, respectively. The residual is $\|\bar{\mathbf{C}}\|_2 = 5.228$ at $N=128$, and no working-graph clique embedding is found at the attempted sizes of $N=176$ and $256$ on Pegasus and $N=128$, $176$, and $256$ on Zephyr, whereas the larger sizes were recorded as \texttt{EMBEDDING\_\allowbreak WALL\_\allowbreak EXCEEDED} without an embedding attempt (see Methods). For the first-order dual scheme of Ohzeki \cite{ohzeki2020breaking}, both the $K=4$-calibrated fixed step (\texttt{Ohzeki} $\eta=0.01 < \eta_{\mathrm{crit}} \approx 0.014\text{--}0.036$) and the $K$-scaled $O(t^{-1/2})$ diminishing-step schedule (\texttt{Ohzeki Dim.}, defined in Methods) avoid the DAC rail saturation of supercritical steps ($\eta=0.05$, which diverges to $269.68\text{--}284.77$ at full chip, see Methods). However, both baselines remain stalled on an $O(1)$ plateau across all ten scales ($1.071\text{--}1.886$ and $1.060\text{--}1.832$ on Pegasus, and $0.885\text{--}2.310$ and $1.015\text{--}1.743$ on Zephyr, respectively). Note that the Zephyr value of $2.310 \pm 2.302$ at $N=64$ (median $1.543$) reflects one stalled seed at $6.734$ and four seeds in the range of $0.377\text{--}2.197$. Conversely, \texttt{Fixed-}$\beta$\texttt{ QSQN} and \texttt{SelfCal-QSQN} reduce the raw QPU ensemble residual substantially below unity across all ten scales, reaching $0.203 \pm 0.032$ at $5{,}612$ qubits ($5.3$-fold and $5.5$-fold lower than $1.071 \pm 0.189$ and $1.120 \pm 0.276$ for \texttt{Ohzeki} $\eta=0.01$ and \texttt{Ohzeki Dim.}, and $0.457 \pm 0.223$ versus $1.612 \pm 0.325$ and $2.246 \pm 0.484$ at $K=8$) and $0.092 \pm 0.038$ at $4{,}560$ qubits ($9.6$-fold and $11.0$-fold lower than $0.885 \pm 0.402$ and $1.015 \pm 0.450$, and $0.270 \pm 0.149$ versus $1.825 \pm 0.364$ and $2.910 \pm 0.522$ at $K=8$). In each of the four full-chip conditions, the final ensemble residual was lower for \texttt{SelfCal-QSQN} than for both calibrated Ohzeki baselines on every one of the six matched instances.

To separate quantum-side dual self-calibration from classical post-processing, the pre-polish raw QPU minimum sample violation $\min_m \|\mathbf{C}(\mathbf{s}_{\mathrm{raw}}^{(m,t)})\|_2$ was extracted directly prior to the 1-bit greedy polish across all trajectories (Fig.~\ref{fig:scaling}b and Table~\ref{tab:scaling}). On full-chip \texttt{Advantage2} ($N=4{,}560, K=4$), \texttt{SelfCal-QSQN} drives the ensemble residual down $58.2$-fold from $5.355 \pm 2.529$ ($t=1$) to $0.092 \pm 0.038$ ($t=30$), while simultaneously reducing the pre-polish raw QPU minimum sample violation $4.9$-fold from $1.190 \pm 0.644$ at $t=1$ ($\boldsymbol{\nu}^{(0)}=\mathbf{0}$) to $0.242 \pm 0.045$ at $t=30$ and reaching a per-seed trajectory minimum of $0.108 \pm 0.017$ (versus $0.155 \pm 0.091$ and $0.182 \pm 0.119$ for \texttt{Ohzeki} $\eta=0.01$ and \texttt{Ohzeki Dim.}, and $0.360 \pm 0.035$ versus $0.798 \pm 0.144$ and $1.399 \pm 0.418$ at $K=8$, respectively). Note that a single spin flip shifts an integer constraint row $A_{\mathrm{raw},ki} \in \{1,\dots,4\}$ by $\pm 2A_{\mathrm{raw},ki}$ ($\ge 2/\sqrt{4{,}560} = 0.0296$ normalized units). Therefore, a pre-polish violation of $0.108$ corresponds to an unnormalized integer $L_2$ residual of $7.31$ units across $4{,}560$ coupled spins. Across $N=64\text{--}4{,}096$ on Zephyr, \texttt{SelfCal-QSQN} maintains pre-polish raw minimum violations of $0.050\text{--}0.140$ ($1.9\text{--}5.5$-fold lower than $0.247\text{--}0.438$ and $0.241\text{--}0.439$ for the two \texttt{Ohzeki} baselines, and $0.305\text{--}2.774$ versus $0.711\text{--}10.436$ and $0.859\text{--}6.374$ for $K=6\text{--}32$ in Fig.~\ref{fig:fdt}b) and yields unpolished zero-violation QPU reads directly on the hardware in $12/30$ runs for $N=32\text{--}512$ (including $4/5$ seeds and $169$ raw feasible reads at $N=64$). Conversely, a single polish pass at $t=1$ ($\boldsymbol{\nu}^{(0)}=\mathbf{0}$, prior to dual calibration) achieves exact feasibility in $0/6$ ($0.0\%$) full-chip Zephyr runs and $1/51$ ($2.0\%$) runs across all ten Zephyr scales. With best-incumbent tracking across all $T$ polish passes of the closed-loop trajectory (see Methods), the 1-bit polish reaches exact zero violation ($\|\mathbf{C}(\mathbf{s}^*)\|_2 = 0.00000$) in $1/5$ ($20\%$) runs at $N=32$ and $4{,}096$, $4/5$ ($80\%$) at $N=64$ and $2{,}048$, $25/25$ ($100\%$) across $N \in \{128, 176, 256, 512, 1024\}$, and $5/6$ ($83.3\%$) at full-chip $N=4{,}560$ (giving $35/41 = 85.4\%$ across the tested Zephyr sizes from 128 spins through full chip, compared with $3/6 = 50.0\%$ and $2/6 = 33.3\%$ at $N=4{,}560$, $0/5$ at $N=4{,}096$, and $21/41 = 51.2\%$ and $22/41 = 53.7\%$ across the corresponding $41$ baseline runs for \texttt{Ohzeki} $\eta=0.01$ and \texttt{Ohzeki Dim.} under the same 1-bit polish procedure and incumbent tracking, with the flip caps and the instance correspondence given in Methods). Note that the full-chip exact-feasible incumbents are reached between iterations $3$ and $28$ (first attained at iterations $3\text{--}16$ across the $5$ feasible seeds), whereas the polish pass of the final iteration alone leaves all $6$ runs at a residual violation of $0.0296$. Across the full-chip tests, none of the three methods returned an exactly feasible raw read. After polish, \texttt{SelfCal-QSQN} reached exact feasibility on one of six Zephyr instances with eight constraints and on one of six and zero of six Pegasus instances with four and eight constraints, respectively.

The size-invariant residual suppression is consistent with Proposition~\ref{prop:fdt} and the $\sqrt{N}$-normalized constraint scaling described in Methods. Inverting the regularized $K \times K$ covariance $\hat{\boldsymbol{\Sigma}}_C^{(t)}$ decouples the overlapping spin subsets across the $K$ constraints while preserving the $1$-to-$1$ native coupler sparsity (i.e., $\mathbf{J}_0$ remains unchanged). Conversely, calibrated subcritical scalar steps ($\eta_t < \eta_{\mathrm{crit}}$) decay quickly along the dominant collective eigenmode of $\boldsymbol{\Sigma}_C$, but stall along the weakly curved differential constraint directions as a result of the Jacobian condition number $\kappa_{\mathrm{cond}} = \lambda_{\max}/\lambda_{\min} \approx 38\text{--}90$, whereas supercritical steps ($\eta_t > \eta_{\mathrm{crit}}$) saturate the $h \in [-4,4]$ DAC rails. In other words, shifting the constraints into second-order preconditioned longitudinal fields eliminates the minor-embedding overhead without sacrificing multi-constraint convergence.

\begin{table*}[!htbp]
  \centering
  \caption{\textbf{Multi-generation native QPU scaling ($K=4$), estimated compute-only runtime, symmetric 1-bit polish baselines, and \texttt{HiGHS} LP bounds.} \texttt{NF} denotes an attempted search that found no clique embedding. \texttt{Skip} denotes a size omitted by the protocol. The abbreviation ref. indicates MILP feasibility on the reference host. Ensemble residuals $\|\bar{\mathbf{C}}\|_2$ report multi-seed means $\pm$ population $\mathrm{s.d.}$ ($\mathrm{ddof}=0$, with $5$ seeds at $S=1{,}000$ for $N \le 4{,}096$ and $6$ seeds at $S=2{,}000$ for full chip). Zephyr columns additionally report the pre-polish trajectory-minimum raw QPU sample violation (\texttt{Raw Min}), the best-incumbent post-polish exact zero-violation rate (\texttt{Exact $0.000$}, where the corresponding rates under the same polish procedure for \texttt{Ohzeki} $\eta=0.01$ / \texttt{Dim.}, which were executed on separately extracted instances for $N \le 4{,}096$ (see Methods), are $1/5, 5/5, 5/5, 2/5, 5/5, 2/5, 3/5, 1/5, 0/5, 3/6$ and $1/5, 4/5, 5/5, 2/5, 4/5, 4/5, 3/5, 2/5, 0/5, 2/6$), pure QPU access time / estimated compute-only time (\texttt{QPU / Comp.\ s}, summing measured QPU access and offline estimates for all $T$ outer-loop CPU updates, and all $T$ 1-bit polish passes, while excluding cloud queue/network latency), exact-feasible objective $f_0(\mathbf{s}^*)$, classical \texttt{HiGHS} ($5.0\text{-s}$ MILP and symmetric LP rounding $+$ 1-bit polish, where \texttt{HiGHS LP + 1-flip} attains $-2.119$ ($4/5$), $-2.858$, $-7.636$, $-7.763$, and $-12.636$ at $N=32, 64, 128, 176,$ and $256$, respectively), and continuous Fortet--McCormick LP lower bound $f_0^{\mathrm{LP}}$. $^{\dagger}$Pegasus $N \le 4{,}096$ used $18$-iteration \texttt{Fixed-}$\beta$\texttt{ QSQN} ($2.23\text{--}4.12\text{ s}$ QPU, $2.24\text{--}4.41\text{ s}$ estimated compute-only), and full-chip $N=5{,}612$ ($30$ iterations, $13.79\text{ s}$ QPU, $14.59\text{ s}$ estimated compute-only) used \texttt{SelfCal-QSQN}. $^{\ddagger}$Averaged over exact-feasible seeds ($-5.884$ matches \texttt{HiGHS} on the feasible $N=32$ seed, and \texttt{HiGHS} achieves $-11.564$ on the $4$ feasible $N=64$ seeds). $^{\S}$Evaluated on the archived $4{,}560$-qubit, $41{,}205$-coupler \texttt{Advantage2\_\allowbreak system1} working graph and averaged over the $5$ exact-feasible seeds ($\mathrm{s.d.}$ $12.6$), where the topology-invariant decoupled LP bound is $-3{,}443.6$.}
  \label{tab:scaling}
  \resizebox{\textwidth}{!}{%
    \begin{tabular}{@{}rrrrrrrrrrrrrr@{}}
      \toprule
      & \multicolumn{4}{c}{\textbf{Pegasus (\texttt{Advantage\_system6}, up to $5{,}612\text{q}$)}} & \multicolumn{6}{c}{\textbf{Zephyr (\texttt{Advantage2\_system1}, up to $4{,}560\text{q}$)}} & \multicolumn{3}{c}{\textbf{Classical MILP, $+1$-Flip \& LP Bound}} \\
      \cmidrule(lr){2-5}\cmidrule(lr){6-11}\cmidrule(lr){12-14}
      \textbf{$N$} & \textbf{Penalty} & \textbf{Ohz.\ ($0.01$)} & \textbf{Ohz.\ (Dim.)} & \textbf{Fixed-$\beta$/SC$^{\dagger}$} & \textbf{Penalty} & \textbf{Ohz.\ ($0.01$)} & \textbf{Ohz.\ (Dim.)} & \textbf{SelfCal $\|\bar{\mathbf{C}}\|_2$} & \textbf{Raw Min / Ex.\ $0$} & \textbf{QPU / Comp.\ (s)} & \textbf{Zeph.\ Feas.\ $f_0$} & \textbf{HiGHS ($+1$-Flip)} & \textbf{LP Bound $f_0^{\mathrm{LP}}$} \\
      \midrule
      32   & $3.624$ ($154\text{q}$)     & $1.450 \pm 0.500$ & $1.454 \pm 0.524$ & $0.439 \pm 0.240$ & $6.534$ ($128\text{q}$) & $1.386 \pm 0.454$ & $1.086 \pm 0.405$ & $0.324 \pm 0.267$ & $0.312$ / $1/5$  & $3.56$ / $3.57$  & $-5.884^{\ddagger}$  & $-5.883$ ($5/5$ opt.) & $-8.828$ \\
      64   & $4.248$ ($436\text{q}$)     & $1.131 \pm 0.362$ & $1.060 \pm 0.179$ & $0.427 \pm 0.154$ & $9.614$ ($415\text{q}$) & $2.310 \pm 2.302$ & $1.204 \pm 0.644$ & $0.144 \pm 0.061$ & $0.050$ / $4/5$  & $3.67$ / $3.68$  & $-15.352^{\ddagger}$ & $-11.788$ ($5/5$, $0$ opt.) & $-24.786$ \\
      128  & $5.228$ ($1{,}640\text{q}$) & $1.203 \pm 0.582$ & $1.447 \pm 0.447$ & $0.458 \pm 0.333$ & \texttt{NF}           & $1.180 \pm 0.442$ & $1.130 \pm 0.432$ & $0.258 \pm 0.159$ & $0.071$ / $5/5$  & $3.83$ / $3.85$  & $-32.569$            & $-24.297$ ($5/5$, $0$ opt.) & $-58.356$ \\
      176  & \texttt{NF}               & $1.701 \pm 0.342$ & $1.607 \pm 0.319$ & $0.330 \pm 0.104$ & \texttt{NF}           & $1.713 \pm 0.713$ & $1.743 \pm 0.727$ & $0.235 \pm 0.090$ & $0.103$ / $5/5$  & $3.92$ / $3.94$  & $-43.912$            & $-31.910$ ($5/5$, $0$ opt.) & $-79.684$ \\
      256  & \texttt{NF}               & $1.534 \pm 0.436$ & $1.505 \pm 0.397$ & $0.300 \pm 0.188$ & \texttt{NF}           & $1.139 \pm 0.663$ & $1.106 \pm 0.680$ & $0.174 \pm 0.112$ & $0.129$ / $5/5$  & $4.04$ / $4.07$  & $-67.638$            & $0/5$ ref. ($-12.636$, $5/5$) & $-125.661$ \\
      512  & \texttt{Skip}               & $1.886 \pm 0.406$ & $1.832 \pm 0.364$ & $0.499 \pm 0.440$ & \texttt{Skip}           & $1.532 \pm 0.484$ & $1.594 \pm 0.500$ & $0.129 \pm 0.078$ & $0.131$ / $5/5$  & $4.23$ / $4.28$  & $-137.97$            & N/A & N/A \\
      1024 & \texttt{Skip}               & $1.297 \pm 0.440$ & $1.335 \pm 0.429$ & $0.575 \pm 0.356$ & \texttt{Skip}           & $1.050 \pm 0.568$ & $1.076 \pm 0.551$ & $0.134 \pm 0.065$ & $0.126$ / $5/5$  & $4.47$ / $4.55$  & $-274.07$            & N/A & N/A \\
      2048 & \texttt{Skip}               & $1.358 \pm 0.465$ & $1.302 \pm 0.522$ & $0.552 \pm 0.334$ & \texttt{Skip}           & $1.315 \pm 0.371$ & $1.347 \pm 0.348$ & $0.120 \pm 0.061$ & $0.139$ / $4/5$  & $4.62$ / $4.78$  & $-545.66^{\ddagger}$ & N/A & N/A \\
      4096 & \texttt{Skip}               & $1.389 \pm 0.380$ & $1.349 \pm 0.371$ & $0.245 \pm 0.067$ & \texttt{Skip}           & $1.345 \pm 0.504$ & $1.389 \pm 0.558$ & $0.163 \pm 0.071$ & $0.140$ / $1/5$  & $4.69$ / $5.08$  & $-1107.7^{\ddagger}$ & N/A & N/A \\
      Full & \texttt{Skip} ($5{,}612\text{q}$) & $1.071 \pm 0.189$ & $1.120 \pm 0.276$ & $\mathbf{0.203 \pm 0.032}$ & \texttt{Skip} ($4{,}560\text{q}$) & $0.885 \pm 0.402$ & $1.015 \pm 0.450$ & $\mathbf{0.092 \pm 0.038}$ & $\mathbf{0.108}$ / $\mathbf{5/6}$ & $12.78$ / $13.44$ & $-1232.7^{\ddagger}$ & $0/6$ ref. ($-13.29$, $1/6$) & $-2607.6^{\S}$ \\
      \bottomrule
    \end{tabular}%
  }
\end{table*}

The estimated compute-only runtime (measured QPU access plus offline estimates for the classical updates and polish) and symmetric 1-bit polish baselines were evaluated against the branch-and-cut \texttt{HiGHS} solver \cite{huangfu2018highs,highs_web} ($5.0\text{-s}$ solver time limit on the Fortet--McCormick linearizations of the identical Zephyr instances, Fig.~\ref{fig:scaling}c and Table~\ref{tab:scaling}). Across $N \in \{32, 64, 128, 176, 256\}$ on Zephyr, the classical outer-loop updates and 1-bit polish passes add $0.007\text{--}0.024\text{ s}$ ($< 0.7\%$), yielding an estimated compute-only runtime of $3.57\text{--}4.07\text{ s}$. At 256 spins, mean QPU access was 4.041 s and the elapsed time recorded by the controller was 44.34 $\pm$ 2.79 s across the five runs. At $N=32$, \texttt{HiGHS} proves optimality on $5/5$ Zephyr seeds ($f_0 = -5.883$) and $5/5$ Pegasus seeds ($-5.872$). \texttt{SelfCal-QSQN} matches the exact optimum ($-5.884$) on its feasible Zephyr seed, and \texttt{Fixed-}$\beta$\texttt{ QSQN} on Pegasus reaches $5/5$ feasibility and matches the global optimum on $3/5$ seeds. For $N=64, 128,$ and $176$ on Zephyr, \texttt{HiGHS} closes $0/5$ optimality gaps within $5.0\text{ s}$ (and applying the 1-bit polish to \texttt{HiGHS} MILP incumbents leaves $f_0$ unchanged since those incumbents already satisfy $\mathbf{C}(\mathbf{s})=\mathbf{0}$), whereas \texttt{SelfCal-QSQN} attains exact-feasible energies lower by $32.8\%$ ($-15.352$ versus $-11.564$ on the $4$ matched seeds at $N=64$), $34.0\%$ ($-32.569$ versus $-24.297$ at $N=128$), and $37.6\%$ ($-43.912$ versus $-31.910$ at $N=176$). At $N=256$, standalone \texttt{HiGHS} MILP returns no integer-feasible solution on any of the $5$ seeds within $5.0\text{ s}$ on the reference host (see Methods). At 256 spins, \texttt{SelfCal-QSQN} and LP marginal rounding followed by the same one-bit polish both recovered feasible solutions on all five instances. The mean feasible energies were $-$67.638 and $-$12.636, respectively. LP rounding and polish took 2.27 s on the reference host using 22 batches of 1,000 draws. Uniform random starts followed by polish gave a mean feasible energy of $-$8.533. This quality gap arises because, on frustrated spin-glass graphs, the continuous Fortet--McCormick LP relaxation collapses spin variables near the unpolarized center $x_i^{\mathrm{LP}} \approx 0.5$ ($s_i^{\mathrm{LP}} \approx 0$) so that auxiliary edge variables absorb bond frustration ($f_0^{\mathrm{LP}} = -125.661$ at $N=256$), and the greedy 1-bit polish projects rounded states onto the nearest feasible hyperplane without descending the rugged two-body energy landscape. At 4,560 spins, \texttt{SelfCal-QSQN} reached exact feasibility after polish on five of six Zephyr instances, with a mean energy of $-$1,232.7 $\pm$ 3.5. The planted states on these five instances had a mean energy of +0.62 $\pm$ 27.73. LP rounding and time-capped polish returned one feasible solution in six runs, at $-$13.29. On the five SelfCal-feasible instances, the mean Fortet--McCormick and decoupled LP lower bounds were $-$2,607.6 $\pm$ 12.6 and $-$3,443.6 $\pm$ 4.0, respectively. Dividing the mean feasible energy by these mean bounds gives 47.3\% and 35.8\%, respectively. The former ratio continues the decreasing trend of $77.0\%$--$53.8\%$ observed on the matched feasible seeds across $N=32\text{--}256$ (with $f_0^{\mathrm{LP}} = -7.641$ and $-24.177$ on the feasible seeds at $N=32$ and $64$, respectively, compared with the $5$-seed means listed in Table~\ref{tab:scaling}), where geometric cycle frustration strictly separates the integer ground state from $f_0^{\mathrm{LP}}$. Note that this fixed-budget evaluation benchmarks general-purpose MILP/LP formulations against the closed-loop hybrid workflow rather than establishing an asymptotic speedup over specialized classical spin-glass heuristics.

\subsection{Fluctuation--dissipation response measurements and self-calibration}

\begin{figure*}[!htbp]
  \centering
  \includegraphics[width=\textwidth]{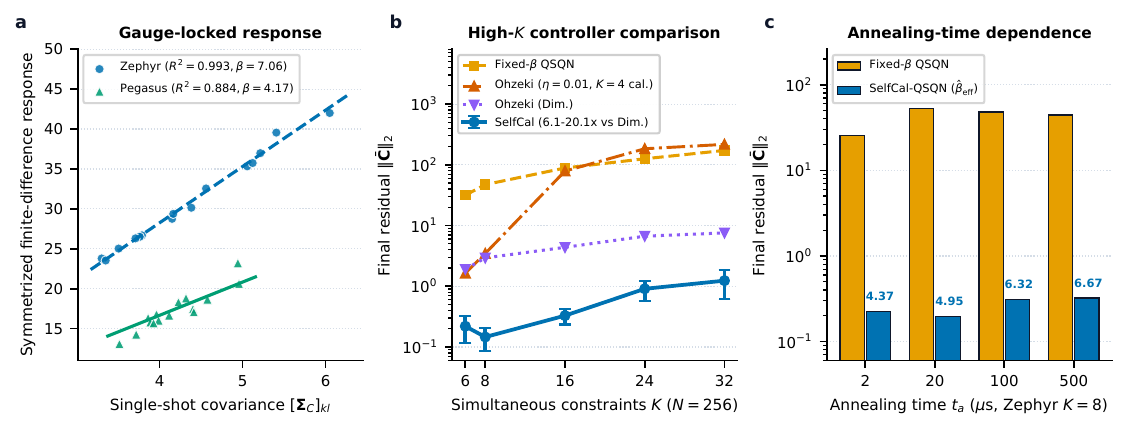}
  \caption{\textbf{Symmetrized hardware response, constraint covariance and controller performance.} \textbf{a}, Gauge-locked ($\mathrm{SRT}=0$, $t=3$) comparison of symmetrized finite-difference response entries $[\mathbf{J}_{\mathrm{FD}}]_{kl}$ against single-shot constraint covariances $[\boldsymbol{\Sigma}_C]_{kl}$ ($N=128, K=5, S=4{,}000$, $15$ unique symmetric elements per processor, with $R^2$ evaluated over the $25$ symmetrized entries as defined in Methods), with zero-intercept reference lines $[\mathbf{J}_{\mathrm{FD}}]_{kl} = \beta_{\mathrm{emp}}[\boldsymbol{\Sigma}_C]_{kl}$ ($\beta_{\mathrm{emp}}=7.06$, Pearson $R^2 = 0.993$, zero-intercept $R_0^2 = 0.992$ on Zephyr, and $\beta_{\mathrm{emp}}=4.17$, Pearson $R^2 = 0.884$, $R_0^2 = 0.811$ on Pegasus, and $R^2 = 0.975, R_0^2 = 0.969$ in the separate $\mathrm{SRT}=4$ probe, see Methods). \textbf{b}, High-$K$ constraint scaling ($K=6\text{--}32$ at $N=256$, $5$ seeds) on Zephyr, comparing \texttt{SelfCal-QSQN} against $K$-scaled diminishing-step \texttt{Ohzeki} (\texttt{Dim.}, $6.1\text{--}20.1\times$ reduction), uncalibrated \texttt{Fixed-}$\beta$\texttt{ QSQN} ($138.4\text{--}323.3\times$ reduction), and fixed-step \texttt{Ohzeki} ($\eta=0.01$ calibrated at $K=4$), where the two \texttt{Ohzeki} baselines were measured on separately extracted instances (see Methods). \textbf{c}, Schedule adaptation across annealing durations $t_a \in \{2, 20, 100, 500\}\,\mu\mathrm{s}$ ($N=256, K=8$, Zephyr). Blue bar labels report the active-descent ($t=2\text{--}12$) median secant gain $\hat{\beta}_{\mathrm{eff}}$ ($4.37\text{--}6.67$), which increases monotonically with $t_a$ together with the mean terminal gain ($12.27\text{--}16.97$). Error bars on \texttt{SelfCal-QSQN} in \textbf{b} denote the standard deviation across seeds (baseline curves plot $5$-seed means for visual clarity).}
  \label{fig:fdt}
\end{figure*}

To compare the hardware response with the covariance predicted by the fluctuation--dissipation theorem (FDT) for the Gibbs model in Proposition~\ref{prop:fdt}, the directional central differences $D_{kl}=[\bar C_k(\boldsymbol{\nu}+\delta\mathbf{e}_l)-\bar C_k(\boldsymbol{\nu}-\delta\mathbf{e}_l)]/(2\delta)$ were measured on both QPUs ($N=128, K=5, S=4{,}000$). Their symmetric part $\mathbf{J}_{\mathrm{FD}}=(\mathbf{D}+\mathbf{D}^{\mathsf T})/2$ was then compared with the single-shot covariance $\boldsymbol{\Sigma}_C$. The probe records retain this symmetric part and do not quantify the antisymmetric response. The measurements were performed at $6$ trajectory checkpoints ($K^2 = 25$ matrix entries and $K(K+1)/2 = 15$ unique symmetric elements per checkpoint) under both the default gauge-locked execution ($\mathrm{SRT}=0$, Fig.~\ref{fig:fdt}a) and randomized gauge averaging ($\mathrm{SRT}=4$, where SRT denotes the number of spin-reversal transforms). Under gauge-locked $\mathrm{SRT}=0$, Zephyr reaches a peak $R^2 = 0.9925$ ($r = 0.9962$ at $t=3$ with $\beta_{\mathrm{emp}}=7.06$, $R^2 = 0.9877$ at $t=14$, and $\beta_{\mathrm{emp}} \in [5.57, 8.17]$ overall), whereas Pegasus yields $\beta_{\mathrm{emp}} \in [3.79, 4.54]$ ($4.17$ and $R^2 = 0.8844$ at $t=3$). In the $\mathrm{SRT}=4$ probes with sub-batch gauge averaging, the linearity reaches $R^2 = 0.9751$ on Pegasus ($\beta_{\mathrm{emp}} \in [3.33, 3.87]$, $3.67$ at $t=10$) and $R^2 = 0.9674$ on Zephyr ($\beta_{\mathrm{emp}} \in [5.58, 7.75]$, $6.56$ at $t=18$, consistent with supplementary $3$-point $\mathrm{SRT}=2$ probes in Methods). Note that the $\mathrm{SRT}=0$ and $\mathrm{SRT}=4$ probes were measured on separately extracted instances with slightly different finite-difference steps (see Methods), and thus the difference between the two settings cannot be attributed to gauge averaging alone. Across these $600$ symmetrized matrix entries ($360$ unique elements across checkpoints, processors and gauge settings), agreement with the covariance varied by checkpoint. Across the six fixed-gauge checkpoints, Pearson $R^2$ over all 25 entries ranged from $0.849$ to $0.993$ on Zephyr and from $0.509$ to $0.884$ on Pegasus. Across the six $\mathrm{SRT}=4$ checkpoints, the corresponding $R^2$ ranged from $0.861$ to $0.967$ on Zephyr and from $0.590$ to $0.975$ on Pegasus. The fitted susceptibility slope was $1.7\text{--}1.8$ times as large on \texttt{Advantage2\_\allowbreak system1} as on Pegasus at the peak-linearity checkpoints ($1.5\text{--}1.9$-fold across active-descent checkpoints). This steeper slope is consistent with the $40\%$ higher Hamiltonian energy scale and lower noise of this processor \cite{dwave_adv2_4400q} and the topology-dependent freeze-out on the $d=20$ Zephyr lattice. At the third gauge-locked iteration, the ten distinct off-diagonal entries (one per constraint pair) gave squared Pearson correlations of 0.994 on Zephyr and 0.933 on Pegasus. Across the six gauge-locked checkpoints, the corresponding squared correlations ranged from 0.837 to 0.994 on Zephyr and from 0.326 to 0.933 on Pegasus. Each response entry averages the two directions of perturbation for that constraint pair. Projecting $\mathbf{J}_{\mathrm{FD}}$ onto the principal eigenvector $\mathbf{v}_{\max}$ of $\mathbf{S}_C$ relative to the $\alpha = 0.30$ shrunk covariance $\boldsymbol{\Sigma}_{\mathrm{shrunk}}$ (see Methods for the definition of $\beta_{\mathrm{dom}}$) yields $\beta_{\mathrm{dom}} \in [7.06, 10.71]$ under $\mathrm{SRT}=0$ on Zephyr ($7.06$ at $t=6$ to $10.71$ at $t=1$, and $[7.13, 10.14]$ under $\mathrm{SRT}=4$), compared with $[5.02, 5.99]$ ($[4.37, 4.99]$ under $\mathrm{SRT}=4$) on Pegasus. Note that even at $N=128, K=5$, the early-descent checkpoints on Zephyr ($\beta_{\mathrm{dom}} = 9.16\text{--}10.71$, corresponding to $0.5\beta_{\mathrm{dom}} = 4.58\text{--}5.36$ within the full $3.53\text{--}5.36$ range) exceed $2\beta_{\mathrm{fixed}} \approx 8.50$, where $\beta_{\mathrm{fixed}} \approx 4.25$ denotes the constant gain of \texttt{Fixed-}$\beta$\texttt{ QSQN} (see Methods). For a local undamped scalar approximation along this mode, $|1-\beta_{\mathrm{dom}}/\beta_{\mathrm{used}}|<1$ requires $\beta_{\mathrm{used}}>\beta_{\mathrm{dom}}/2$. The early $K=5$ measurements therefore suggest that a fixed gain of $4.25$ could produce an excessively large update. The damped trajectories at $K\geq6$ were measured separately, without a corresponding dominant-mode response probe.

Fig.~\ref{fig:fdt}b,c shows the performance of the zero-overhead SNR-gated secant estimator $\hat{\beta}_{\mathrm{eff}}^{(t)}$ for high constraint counts $K \in \{6, 8, 16, 24, 32\}$ ($N=256$, $5$ seeds) and annealing times $t_a \in \{2, 20, 100, 500\}\,\mu\mathrm{s}$ ($N=256, K=8$) \cite{dwave_annealing_doc}. In \texttt{Fixed-}$\beta$\texttt{ QSQN}, the susceptibility gain is fixed to the hardware-agnostic analytical constraint-variance prior $\beta_{\mathrm{fixed}} = ((N/32)\sigma_A^2)^{-1} \approx 4.25$, where $\sigma_A^2$ denotes the mean squared entry of $\mathbf{A}$ (see Methods). On Pegasus, this prior lies near the measured susceptibility scale. The fixed-gain controller yielded residuals of $\|\bar{\mathbf{C}}\|_2=0.412$ at $K=8$ and $1.690$ at $K=32$. On Zephyr, the high-$K$ fixed-gain residual trajectories exhibited oscillations. This behavior is consistent with the large-step interpretation suggested by the $K=5$ probes. The damped high-$K$ runs do not establish its cause. In addition, the high-$K$ scaling exposes the step-size trade-off of open-loop first-order dual methods, which update $\boldsymbol{\nu}^{(t)}$ without closed-loop overshoot backtracking (see Methods). Note that the calibrated \texttt{Ohzeki} baselines of this benchmark were executed on separately extracted instances of the same generator, sizes, and seeds (see Methods). At $K=6\text{--}8$, the fixed step $\eta=0.01$ calibrated at $K=4$ yielded moderate residuals ($1.641\text{--}3.472$ on Zephyr and $1.574\text{--}1.837$ on Pegasus). At $K\geq16$, the same step yielded substantially larger residuals ($78.90 \pm 19.88$, $183.95 \pm 3.59$, and $217.14 \pm 6.30$ at $K=16, 24,$ and $32$ on Zephyr, and $51.66\text{--}209.63$ on Pegasus). Scaling the step inversely with $K$ in \texttt{Ohzeki Dim.} prevents the divergence at $K \ge 16$. However, the residual stalls across $K=6\text{--}32$ at $1.908\text{--}7.572$ on Zephyr ($4.380\text{--}7.572$ for $K \ge 16$) and $2.002\text{--}7.138$ on Pegasus. Conversely, \texttt{SelfCal-QSQN} converges to $0.146 \pm 0.061$ at $K=8$ on Zephyr ($20.1$-fold lower than $2.938$ for \texttt{Ohzeki Dim.}, $323.3$-fold lower than $47.23$ for \texttt{Fixed-}$\beta$\texttt{ QSQN}, and $23.8$-fold lower than $3.472$ for $K=4$-calibrated \texttt{Ohzeki} $\eta=0.01$) and to $0.329 \pm 0.094$ at $K=16$ ($13.3$-fold lower than $4.380$ for \texttt{Ohzeki Dim.} and $239.8$-fold lower than unguarded \texttt{Ohzeki} $\eta=0.01$). In addition, \texttt{SelfCal-QSQN} converges to $0.909 \pm 0.327$ at $K=24$ ($7.3$-fold lower than $6.681$ for \texttt{Ohzeki Dim.} and $138.4$-fold lower than $125.79$ for \texttt{Fixed-}$\beta$\texttt{ QSQN}) and to $1.238 \pm 0.625$ at $K=32$ ($6.1$-fold lower than $7.572$ and $139.4$-fold lower than $172.53$, respectively, and $0.936 \pm 0.258$ versus $7.138 \pm 0.681$ on Pegasus). Note that the overshoot safeguard fired in $0/10$ Zephyr runs at $K=6\text{--}8$ and activated at $K \ge 16$. Therefore, the $K=6\text{--}8$ gains reflect covariance shrinkage, the self-calibrated susceptibility gain (a topology-level prior refined by the secant updates), and local-field trust-region scaling without safeguard intervention, whereas the $K \ge 16$ gains reflect their combined action with overshoot safeguarding. Over $t_a = 2\,\mu\mathrm{s}$ to $500\,\mu\mathrm{s}$ on Zephyr ($K=8$, Fig.~\ref{fig:fdt}c), \texttt{Fixed-}$\beta$\texttt{ QSQN} degrades to $25.51\text{--}52.94$, whereas \texttt{SelfCal-QSQN} maintains $\|\bar{\mathbf{C}}\|_2 \in [0.196, 0.322]$. Across this $250$-fold schedule sweep, the active-descent ($t=2\text{--}12$) median secant estimate rises monotonically from $4.37$ to $6.67$ ($4.37 \to 4.95 \to 6.32 \to 6.67$), which is consistent with slow-anneal freeze-out cooling. In addition, the late-iteration positive SNR-gating selection at small step norms ($\|\Delta\boldsymbol{\nu}^{(t-1)}\|_2 \ll 1$, supplemented in $1/20$ schedule runs and at $K \ge 16$ by overshoot-safeguard boosts, see Methods) elevates the mean terminal controller gain $\hat{\beta}_{\mathrm{eff}}^{(T)}$ from $12.27$ to $16.97$. This result indicates that $\hat{\beta}_{\mathrm{eff}}^{(T)}$ acts as an effective closed-loop damping gain rather than a direct physical thermometer. Accordingly, the online secant self-calibration was enabled by default in all of the subsequent benchmarks without per-instance tuning of $\beta$. The 6.1--20.1-fold reductions relative to diminishing steps compare group means on separately generated instances. The controller comparison changes the initial gain, step limiting and safeguards together with secant calibration, so it does not isolate the contribution of the secant update.

\subsection{Impact of shot noise and reverse annealing}

\begin{figure*}[!htbp]
  \centering
  \includegraphics[width=\textwidth]{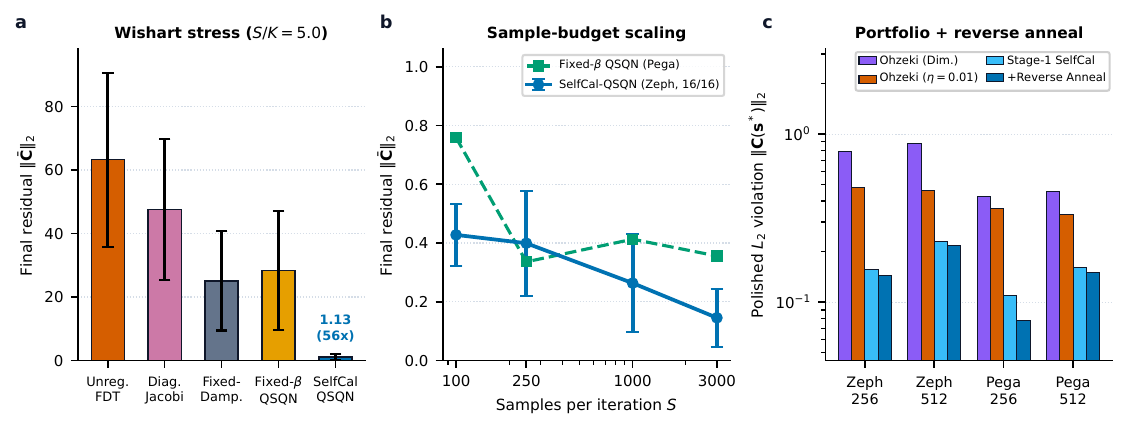}
  \caption{\textbf{Progressive Wishart stabilization, sample-budget scaling, and two-stage reverse-annealing portfolio optimization.} \textbf{a}, Progressive five-variant stabilization benchmark under extreme finite-sample shot noise ($N=128, K=12, S=60$, giving $S/K=5.0$ across $10$ seeds on Zephyr), showing a $56.1\times$ combined error suppression relative to \texttt{Unreg.\ FDT}. \textbf{b}, Sample-budget scaling for $S \in \{100, 250, 1000, 3000\}$ ($N=256, K=4$), comparing \texttt{SelfCal-QSQN} on Zephyr ($16/16$ exact zero polished violation across all $S \ge 100$) with \texttt{Fixed-}$\beta$\texttt{ QSQN} on Pegasus. \textbf{c}, Constrained multi-sector portfolio optimization ($N \in \{256, 512\}, K=6$, $8$ seeds per bar), comparing 1-bit polished $L_2$ violations $\|\mathbf{C}(\mathbf{s}^*)\|_2$ across \texttt{Ohzeki} (\texttt{Dim.} and $\eta=0.01$), Stage-1 \texttt{SelfCal-QSQN}, and Stage-2 paused reverse annealing at locked $\boldsymbol{\nu}^*$, where the \texttt{Ohzeki} baselines were measured on separately extracted instances (see Methods). Error bars in \textbf{a} ($10$ seeds) and on \texttt{SelfCal-QSQN} in \textbf{b} ($4$ seeds per point) denote the standard deviation across seeds (the Pegasus \texttt{Fixed-}$\beta$ curve in \textbf{b} plots $4$-seed means for clarity).}
  \label{fig:ablation}
\end{figure*}

To evaluate cumulative controller stabilization under severe quantum shot noise, five progressive curvature variants were benchmarked at $N=128, K=12$ using only $S=60$ reads per iteration ($S/K = 5.0$, $10$ seeds per QPU, Fig.~\ref{fig:ablation}a and Methods). On Zephyr, unregularized FDT covariance inversion (\texttt{Unreg.\ FDT}) diverges to $\|\bar{\mathbf{C}}\|_2 = 63.17 \pm 27.42$, and discarding off-diagonal couplings (\texttt{Diag.\ Jacobi}) stalls at $47.48 \pm 22.22$. Retaining off-diagonal covariances via shrinkage under fixed damping (\texttt{Fixed-Damp.}) lowers the error $1.9$-fold relative to \texttt{Diag.\ Jacobi} ($2.5$-fold relative to \texttt{Unreg.\ FDT}) to $25.04 \pm 15.60$, whereas adding residual-adaptive damping at fixed $\beta_{\mathrm{fixed}}$ (\texttt{Fixed-}$\beta$\texttt{ QSQN}) remains stalled at $28.44 \pm 18.71$ on Zephyr (versus $1.640 \pm 0.760$ on Pegasus). Combining diagonal-target shrinkage (equation~\eqref{eq:shrink}) with EMA covariance memory, online secant $\hat{\beta}_{\mathrm{eff}}^{(t)}$ self-calibration, non-monotone overshoot safeguarding, and direction-preserving local-field trust-region scaling in \texttt{SelfCal-QSQN} reduces the Zephyr residual $56.1$-fold overall relative to \texttt{Unreg.\ FDT} to $1.127 \pm 0.922$ (median $0.713$), and reduces the Pegasus standard deviation $2.5$-fold relative to \texttt{Fixed-Damp.} ($1.088 \pm 0.334$ versus $1.175 \pm 0.847$). Note that because the transition from \texttt{Fixed-}$\beta$\texttt{ QSQN} to \texttt{SelfCal-QSQN} jointly activates EMA memory, secant calibration, overshoot safeguarding, and field trust-region scaling (together with an increase in the shrinkage weight from $0.30$ to $0.35$), the $56.1$-fold reduction reflects their combined stabilization under Wishart noise rather than any single component in isolation.

When the per-iteration read count is swept over $S \in \{100, 250, 1000, 3000\}$ on Zephyr ($N=256, K=4$, Fig.~\ref{fig:ablation}b), the ensemble residual decays monotonically ($0.428 \pm 0.105 \to 0.399 \pm 0.179 \to 0.264 \pm 0.166 \to 0.146 \pm 0.098$), and exact $0.0000$ polished feasibility is achieved in $16/16$ ($100\%$) of the runs, even at $S=100$. The spin-reversal comparison \cite{dwave_solver_doc} used $S=800$ reads per iteration and four seeds per setting. The gauge-locked execution ($\mathrm{SRT}=0$) achieves an ensemble residual of $0.126 \pm 0.080$, whereas the randomized sub-batch gauge averaging ($\mathrm{SRT}=4$) yields $0.512 \pm 0.453$. All four runs in each setting reached exact feasibility after the one-bit polish.

The two-stage optimization of a multi-sector portfolio was evaluated next ($N \in \{256, 512\}$, $K=6$ coupled equality constraints comprising one cardinality constraint, four sector capital quotas, and one market-beta exposure constraint, $8$ seeds per bar, Fig.~\ref{fig:ablation}c). In Stage~1, the calibrated first-order baselines (executed on separately extracted instances of the same generator, see Methods) stall at ensemble residuals of $6.62, 9.50, 7.74,$ and $10.16$ for \texttt{Ohzeki} ($\eta=0.01$) and $8.42, 11.81, 9.84,$ and $12.52$ for \texttt{Ohzeki Dim.} (for Pegasus $N=256, 512$ and Zephyr $N=256, 512$, respectively), whereas Stage-1 \texttt{SelfCal-QSQN} converges to $0.613, 0.966, 0.730,$ and $0.976$, respectively ($9.8\text{--}10.8$-fold lower than \texttt{Ohzeki} $\eta=0.01$ and $12.2\text{--}13.7$-fold lower than \texttt{Ohzeki Dim.}). Once Stage~1 locks $\boldsymbol{\nu}^*$, Stage-2 reverse annealing with a pause \cite{marshall2019power,chen2020pausing} ($s_{\mathrm{pause}} \in \{0.48, 0.58\}$, $15\,\mu\mathrm{s}$ dwell) uses the two best Stage-1 states as initial configurations under $\mathbf{h}^{\mathrm{eff}}(\boldsymbol{\nu}^*)$ prior to the identical 1-bit greedy polish applied to all methods. Evaluated on polished $L_2$ violations $\|\mathbf{C}(\mathbf{s}^*)\|_2$ (Fig.~\ref{fig:ablation}c), Stage~2 on Pegasus ($N=256$) lowers the Stage-1 violation by $28.6\%$ ($0.1094 \to 0.0781$, reaching exact $0.0000$ violation on $4/8$ seeds versus $0/8$ for both \texttt{Ohzeki} baselines, i.e., $4.6$-fold and $5.4$-fold lower than $0.3614$ and $0.4244$, respectively). On Zephyr ($N=512$), Stage~2 lowers the mean polished portfolio energy from $-65.53$ to $-66.57$ ($\Delta f_0 = -1.034$), which is substantially lower than $-28.50$ for \texttt{Ohzeki} $\eta=0.01$, $-27.20$ for \texttt{Ohzeki Dim.}, and $+46.65$ for the planted reference, while reducing the polished violation to $0.2180$ ($2.1$-fold lower than $0.4609$ for \texttt{Ohzeki} $\eta=0.01$ and $4.0$-fold lower than $0.8799$ for \texttt{Ohzeki Dim.}). Note that the Stage-2 incumbent is initialized with the best polished Stage-1 state, and thus the Stage-2 violation cannot exceed the Stage-1 violation by construction. The improvement was measured after reverse annealing and polish. The experiment included no forward-annealing control with the same additional read budget.

\subsection{Non-native graphs and mixed inequality constraints}

\begin{figure*}[!htbp]
  \centering
  \includegraphics[width=\textwidth]{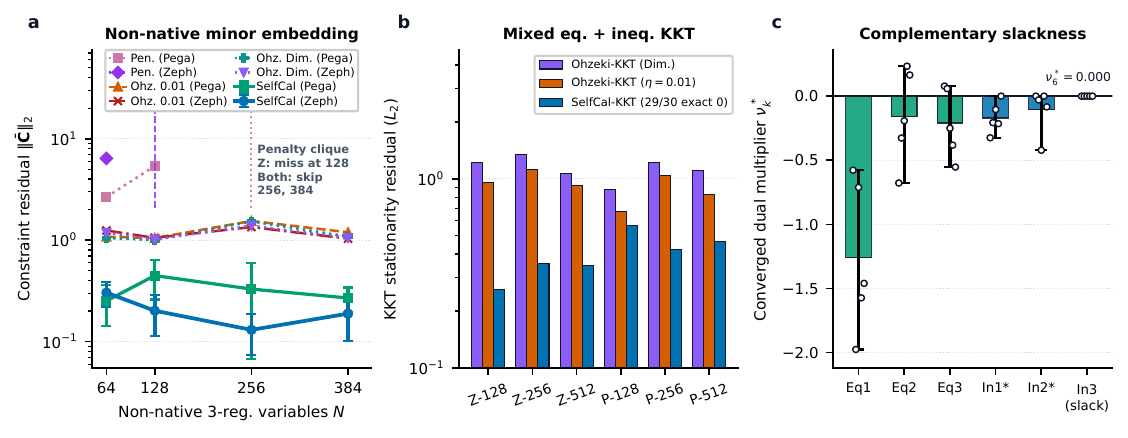}
  \caption{\textbf{Minor-embedded non-native sparse graphs and slack-free mixed KKT optimization.} \textbf{a}, Non-native $3$-regular/modular sparse graphs ($N=64\text{--}384, K=4$, $5$ seeds per processor). \texttt{SelfCal-QSQN} embeds only the sparse $\mathbf{J}_0$ (chain length $1.11\text{--}1.45$) and attains exact zero polished violation in $38/40$ ($95.0\%$) of the runs ($100\%$ for all $N \ge 128$) against both calibrated \texttt{Ohzeki} baselines on Pegasus and Zephyr. \textbf{b}, Mixed equality and inequality KKT stationarity residuals ($K_{\mathrm{eq}}=3, K_{\mathrm{ineq}}=3$, zero slack qubits, $N=128\text{--}512$), with exact mixed feasibility in $29/30$ ($96.7\%$) of the runs (the \texttt{Ohzeki-KKT} baselines were measured on separately extracted instances, see Methods). \textbf{c}, Converged dual multipliers $\boldsymbol{\nu}^*$ on Zephyr ($N=256$, bars show $5$-seed means, whiskers and markers show $5$-seed min--max ranges), which confirm non-positive dual-cone enforcement ($\nu_4^*, \nu_5^* \le 0$) on the tight capacity caps (\texttt{In1*}, \texttt{In2*}) and exact complementary slackness ($\nu_6^* = 0.00000$) on the slack capacity cap (\texttt{In3}). Error bars on the \texttt{SelfCal-QSQN} curves in \textbf{a} denote the standard deviation across seeds (baseline curves plot $5$-seed means for visual clarity).}
  \label{fig:phase3}
\end{figure*}

Real-world optimization graphs do not necessarily match the Pegasus or Zephyr topology natively. Therefore, the proposed scheme was evaluated on non-native $3$-regular plus modular non-planar sparse graphs ($N \in \{64, 128, 256, 384\}, K=4$, $40$ QPU runs, Fig.~\ref{fig:phase3}a and Table~\ref{tab:phase3}). The quadratic-penalty QUBO scheme requires $1{,}640$ physical qubits (mean chain length $12.81$) at $N=128$ on Pegasus. The non-native benchmark found no penalty-clique embedding at 128 spins on Zephyr. The protocol skipped the 256- and 384-spin penalty searches on both processors. Conversely, \texttt{SelfCal-QSQN} minor-embeds only the intrinsic sparse graph $\mathbf{J}_0$ once, and uses a mean of $186$ qubits at $N=128$ on Pegasus ($8.8$-fold fewer qubits, chain length $1.45$). Across $N=64\text{--}384$ with compact static chains ($1.11\text{--}1.45$), \texttt{SelfCal-QSQN} reduces ensemble residuals $2.2\text{--}4.7$-fold on Pegasus ($0.252\text{--}0.448$ versus $1.050\text{--}1.548$ for \texttt{Ohzeki} $\eta=0.01$ and $1.001\text{--}1.531$ for \texttt{Ohzeki Dim.}) and $3.9\text{--}10.6$-fold on Zephyr ($0.131\text{--}0.304$ versus $1.035\text{--}1.347$ and $1.016\text{--}1.391$). In addition, exact $0.0000$ polished feasibility is achieved in $38/40$ ($95.0\%$) of the runs ($30/30 = 100\%$ for all $N \ge 128$). This is because the linear distribution of the dual field shifts across the short static chains ($h_q^{\mathrm{phys},(t)} = w_{i,q} h_i^{\mathrm{eff}}(\boldsymbol{\nu}^{(t)})$) avoids both the $O(N^2)$ clique chains and the per-iteration re-embedding overhead.

Finally, Fig.~\ref{fig:phase3}b,c and Table~\ref{tab:phase3} present the results obtained for the slack-free mixed equality and inequality benchmarks ($N \in \{128, 256, 512\}$, $30$ QPU runs, $0$ slack qubits). In these benchmarks, $K_{\mathrm{eq}}=3$ equalities are combined with $K_{\mathrm{ineq}}=3$ inequalities comprising two tight capacity caps and one slack capacity cap. The active-set \texttt{SelfCal-QSQN-KKT} scheme reduces the KKT stationarity residuals $\|\mathbf{r}_{\mathrm{KKT}}\|_2$ (the natural KKT residual defined in Methods) to $0.260\text{--}0.355$ on Zephyr ($2.7\text{--}3.7$-fold lower than $0.925\text{--}1.114$ for \texttt{Ohzeki-KKT} $\eta=0.01$ and $3.1\text{--}4.7$-fold lower than $1.064\text{--}1.344$ for \texttt{Ohzeki-KKT Dim.}) and $0.422\text{--}0.569$ on Pegasus (versus $0.675\text{--}1.041$ and $0.876\text{--}1.220$). Note that the \texttt{Ohzeki-KKT} baselines of the mixed benchmarks were executed on separately extracted instances of the same generator, sizes, and seeds, whereas the non-native baselines above were executed on the identical instances (see Methods). In addition, exact $0.0000$ mixed primal feasibility is achieved in $29/30$ ($96.7\%$) of the runs ($15/15$ on Pegasus and $14/15$ on Zephyr, compared with $26/30 = 86.7\%$ for \texttt{Ohzeki-KKT} $\eta=0.01$ and $24/30 = 80.0\%$ for \texttt{Ohzeki-KKT Dim.}). In accordance with Case~2 in the proof of Proposition~\ref{prop:fdt}, the converged multipliers in Fig.~\ref{fig:phase3}c satisfy strict KKT complementary slackness. Specifically, the two tight capacity caps remain within the non-positive dual cone ($\nu_4^*, \nu_5^* \le 0$, taking negative values on every seed where the cap binds and zero otherwise), whereas the slack capacity cap converges identically to $\nu_6^* = 0.00000$ in all $30$ runs without consuming any slack qubits.

\begin{table*}[!htbp]
  \centering
  \caption{\textbf{Non-native minor-embedded sparse graph scaling and slack-free mixed equality--inequality KKT benchmarks.} \texttt{NF} denotes an unsuccessful search and \texttt{Skip} a search omitted by the protocol. Summary of the Phase-3 and Phase-4 QPU runs ($5$ seeds per entry, reporting $5$-seed means $\pm$ population $\mathrm{s.d.}$, $\mathrm{ddof}=0$, where the Phase-4 baselines share the identical instances for the non-native graphs and use separately extracted instances for the mixed KKT benchmarks, see Methods) comparing quadratic-penalty QUBO, $K=4$-calibrated fixed-step \texttt{Ohzeki} ($\eta=0.01$), $K$-scaled $O(t^{-1/2})$ diminishing-step \texttt{Ohzeki} (\texttt{Dim.}), and \texttt{SelfCal-QSQN} on \texttt{Advantage\_system6} (Pegasus) and \texttt{Advantage2\_system1} (Zephyr).}
  \label{tab:phase3}
  \resizebox{\textwidth}{!}{%
    \begin{tabular}{@{}llrrrrrrrr@{}}
      \toprule
      \textbf{Benchmark Suite} & \textbf{Processor} & \textbf{$N$} & \textbf{Phys. Qubits (SelfCal vs Pen.)} & \textbf{Mean Chain} & \textbf{Penalty QUBO} & \textbf{Ohzeki ($\eta=0.01$)} & \textbf{Ohzeki (Dim.)} & \textbf{SelfCal-QSQN Residual} & \textbf{Exact $0.000$ Rate} \\
      \midrule
      Non-Native 3-Reg. ($K=4$) & Pegasus & 64  & $77\text{q}$ vs $436\text{q}$ ($5.7\times$)     & $1.203$ & $2.663$  & $1.082 \pm 0.540$ & $1.053 \pm 0.534$ & $0.252 \pm 0.109$ & $5/5$ ($100\%$) \\
      Non-Native 3-Reg. ($K=4$) & Pegasus & 128 & $186\text{q}$ vs $1{,}640\text{q}$ ($8.8\times$) & $1.452$ & $5.384$  & $1.050 \pm 0.079$ & $1.001 \pm 0.077$ & $0.448 \pm 0.188$ & $5/5$ ($100\%$) \\
      Non-Native 3-Reg. ($K=4$) & Pegasus & 256 & $304\text{q}$ vs \texttt{Skip} ($32{,}640$ cpl.) & $1.188$ & \texttt{Skip} & $1.548 \pm 0.209$ & $1.531 \pm 0.217$ & $0.330 \pm 0.262$ & $5/5$ ($100\%$) \\
      Non-Native 3-Reg. ($K=4$) & Pegasus & 384 & $477\text{q}$ vs \texttt{Skip} ($73{,}536$ cpl.) & $1.243$ & \texttt{Skip} & $1.197 \pm 0.328$ & $1.087 \pm 0.356$ & $0.269 \pm 0.072$ & $5/5$ ($100\%$) \\
      Non-Native 3-Reg. ($K=4$) & Zephyr  & 64  & $71\text{q}$ vs $415\text{q}$ ($5.8\times$)     & $1.112$ & $6.405$  & $1.247 \pm 0.663$ & $1.188 \pm 0.623$ & $0.304 \pm 0.084$ & $3/5$ ($60\%$)  \\
      Non-Native 3-Reg. ($K=4$) & Zephyr  & 128 & $163\text{q}$ vs \texttt{NF} ($8{,}128$ cpl.)  & $1.277$ & \texttt{NF} & $1.054 \pm 0.331$ & $1.016 \pm 0.320$ & $0.202 \pm 0.088$ & $5/5$ ($100\%$) \\
      Non-Native 3-Reg. ($K=4$) & Zephyr  & 256 & $283\text{q}$ vs \texttt{Skip} ($32{,}640$ cpl.) & $1.106$ & \texttt{Skip} & $1.347 \pm 0.339$ & $1.391 \pm 0.317$ & $0.131 \pm 0.057$ & $5/5$ ($100\%$) \\
      Non-Native 3-Reg. ($K=4$) & Zephyr  & 384 & $435\text{q}$ vs \texttt{Skip} ($73{,}536$ cpl.) & $1.133$ & \texttt{Skip} & $1.035 \pm 0.419$ & $1.055 \pm 0.398$ & $0.189 \pm 0.086$ & $5/5$ ($100\%$) \\
      \midrule
      Mixed KKT ($3\text{ eq}+3\text{ ineq}$) & Pegasus & 128 / 256 / 512 & Native 1-to-1 ($0$ slack q.) & $1.000$ & --- & $0.675 / 1.041 / 0.830$ & $0.876 / 1.220 / 1.111$ & $0.569 / 0.422 / 0.465$ & $\mathbf{15/15}$ ($\mathbf{100\%}$) \\
      Mixed KKT ($3\text{ eq}+3\text{ ineq}$) & Zephyr  & 128 / 256 / 512 & Native 1-to-1 ($0$ slack q.) & $1.000$ & --- & $0.952 / 1.114 / 0.925$ & $1.225 / 1.344 / 1.064$ & $0.260 / 0.355 / 0.347$ & $\mathbf{14/15}$ ($\mathbf{93.3\%}$) \\
      \bottomrule
    \end{tabular}%
  }
\end{table*}

\section{Discussion}

This study has proposed an accurate, robust, and scalable hardware-native quantum--classical closed-loop workflow, designated as \texttt{SelfCal-QSQN}, for constrained Ising optimization on sparse superconducting processors with zero auxiliary quadratic penalty couplings. In the proposed workflow, linear equalities and inequalities are shifted strictly into the longitudinal local fields, and the regularized susceptibility curvature is reconstructed from single-shot spin fluctuations. As a result, finite-temperature readout variance is converted into a well-conditioned second-order preconditioner across thousands of physical qubits without modifying the native coupler graph. Encoding the tested linear constraints in the local fields avoids the dense couplings introduced by squaring them. The need to embed the objective still depends on its graph.

Several operational boundaries and practical trade-offs of the proposed workflow are discussed in the following. First, pure longitudinal local-field dualization applies specifically to constraints that are linear in $\mathbf{s}$, since non-linear constraint terms alter the pairwise spin couplings unless their interaction support already lies within $\mathcal{E}$. Second, excluding cloud queue and SAPI network latency, the estimated compute-only runtime on Zephyr is $3.57\text{--}5.08\text{ s}$ per $22$-iteration run across $N=32\text{--}4{,}096$ ($3.56\text{--}4.69\text{ s}$ pure QPU access time, $0.004\text{--}0.11\text{ s}$ classical dual updates, and $0.003\text{--}0.28\text{ s}$ across all $22$ 1-bit polish passes) and $13.44\text{ s}$ at full-chip $N=4{,}560$ ($12.78\text{ s}$ pure QPU access, $0.31\text{ s}$ dual updates, and $0.35\text{ s}$ for $30$ polish passes), with $14.59\text{ s}$ at $5{,}612$ qubits on Pegasus \cite{dwave_timing_doc}. Thus, classical outer-loop and polish computations account for less than $0.7\%$ of estimated compute-only time for $N \le 256$ and $4.9\%\text{--}5.5\%$ at full chip. The archived controller elapsed times also include host and service delays. The records do not resolve the individual contributions. Third, a small ensemble residual can coexist with substantial violations in individual reads. The full-chip runs yielded no exactly feasible raw read. Accordingly, exact feasibility at full chip is obtained by means of best-incumbent tracking across the polish passes rather than from the final iteration alone. Conversely, the symmetric 1-bit polish ablations show that a single polish pass on uncalibrated QPU readouts reaches exact feasibility in $1/51$ runs, and that the polish yields high-energy uncorrelated states when applied to fractional LP relaxations. Therefore, low objective energy and exact hard-constraint satisfaction arise jointly from susceptibility-preconditioned quantum sampling and lightweight local refinement, and the fixed-budget MILP comparisons show the topological scalability and practical feasibility convergence of the hybrid workflow rather than a standalone asymptotic quantum speedup.

Beyond superconducting quantum annealers, the mathematical structure of \texttt{SelfCal-QSQN} depends only on access to finite spin samples from a parameterized Ising energy model. Therefore, the same shrinkage-stabilized, secant-calibrated dual controller is directly compatible with coherent optical Ising machines \cite{inagaki2016cim}, neutral-atom Rydberg arrays \cite{ebadi2022rydberg}, trapped-ion simulators, and digital CMOS or bifurcation annealers \cite{aramon2019da,goto2019sb,mohseni2022ising}. In other words, the proposed workflow provides a hardware-native foundation for large-scale linearly constrained discrete optimization across physical Ising platforms. The future work will investigate co-located low-latency controller integration and the deployment of the proposed workflow on these alternative physical Ising machines.

\section{Methods}

\subsection{Proof of Proposition 1}

\noindent\textbf{Proof:} The exact fluctuation--dissipation gradient and Hessian identities for Problem~(P2) are proven first. Let $Z(\boldsymbol{\nu}) = \sum_{\mathbf{s} \in \{-1,+1\}^N} \exp[-\beta_{\mathrm{eff}}H_{\mathrm{eff}}(\mathbf{s};\boldsymbol{\nu})]$ denote the partition function at the effective freeze-out inverse temperature $\beta_{\mathrm{eff}} > 0$. Since $\partial H_{\mathrm{eff}}(\mathbf{s};\boldsymbol{\nu})/\partial\nu_k = -(\mathbf{a}_k^{\mathsf{T}}\mathbf{s} - B_k) = -C_k(\mathbf{s})$, differentiating $\Phi(\boldsymbol{\nu}) = -\beta_{\mathrm{eff}}^{-1}\ln Z(\boldsymbol{\nu})$ with respect to $\nu_k$ yields $\partial\Phi/\partial\nu_k = -Z(\boldsymbol{\nu})^{-1}\sum_{\mathbf{s}} C_k(\mathbf{s})\exp[-\beta_{\mathrm{eff}}H_{\mathrm{eff}}(\mathbf{s};\boldsymbol{\nu})] = -\bar{C}_k(\boldsymbol{\nu})$. Differentiating a second time with respect to $\nu_l$ gives
\begin{equation}
  \frac{\partial^2 \Phi(\boldsymbol{\nu})}{\partial\nu_k\,\partial\nu_l} = -\frac{\partial\bar{C}_k(\boldsymbol{\nu})}{\partial\nu_l} = -\beta_{\mathrm{eff}}\Bigl(\langle C_k C_l\rangle_{\boldsymbol{\nu}} - \bar{C}_k(\boldsymbol{\nu})\bar{C}_l(\boldsymbol{\nu})\Bigr),
  \label{eq:fdt}
\end{equation}
where $[\boldsymbol{\Sigma}_C(\boldsymbol{\nu})]_{kl} = \langle C_k C_l\rangle_{\boldsymbol{\nu}} - \bar{C}_k\bar{C}_l$ forms a symmetric positive semi-definite covariance matrix. Therefore, $\nabla_{\boldsymbol{\nu}}^2\Phi(\boldsymbol{\nu}) = -\beta_{\mathrm{eff}}\boldsymbol{\Sigma}_C(\boldsymbol{\nu}) \preceq \mathbf{0}$ and $\nabla_{\boldsymbol{\nu}}\bar{\mathbf{C}}(\boldsymbol{\nu}) = +\beta_{\mathrm{eff}}\boldsymbol{\Sigma}_C(\boldsymbol{\nu}) \succeq \mathbf{0}$.

The positive definiteness, active-subspace ascent, and dual-space equivalence of the regularized curvature operator $\mathbf{M}_t = \kappa\hat{\beta}_{\mathrm{eff}}^{(t)}\hat{\boldsymbol{\Sigma}}_C^{(t)} + \lambda_t\mathbf{I}_K$ with $\kappa > 0$, $\hat{\beta}_{\mathrm{eff}}^{(t)} > 0$, and $\lambda_t > 0$ are established next. Denoting the equality and inequality row index sets by $\mathcal{I}_{\mathrm{eq}} = \{1,\dots,K_{\mathrm{eq}}\}$ and $\mathcal{I}_{\mathrm{ineq}} = \{K_{\mathrm{eq}}+1,\dots,K\}$, two cases are considered, namely Case~1 (equality rows $k \in \mathcal{I}_{\mathrm{eq}}$ and active inequality rows $k \in \mathcal{A}_t \subseteq \mathcal{I}_{\mathrm{ineq}}$) and Case~2 (inactive KKT inequality rows $k \in \mathcal{I}_{\mathrm{ineq}}\setminus\mathcal{A}_t$). In Case~1, $\hat{\boldsymbol{\Sigma}}_C^{(t)} \succeq \mathbf{0}$ is a convex combination of positive semi-definite sample covariances and their positive semi-definite diagonal parts, and thus every eigenvalue of $\mathbf{M}_t$ satisfies $\lambda_{\min}(\mathbf{M}_t) \ge \lambda_t > 0$. Thus, $\mathbf{M}_t$ is strictly positive definite. Furthermore, for any nonzero active residual $[\mathbf{g}_{\mathrm{eff}}^{(t)}]_{\mathcal{W}} \ne \mathbf{0}$ ($\mathcal{W} = \mathcal{I}_{\mathrm{eq}} \cup \mathcal{A}_t$), the inner product with the sampled estimate $-\mathbf{g}_{\mathrm{eff}}^{(t)}$ of the dual free-energy gradient satisfies $(-\mathbf{g}_{\mathrm{eff}}^{(t)})^{\mathsf{T}}\Delta\boldsymbol{\nu}_{\mathrm{raw}}^{(t)} = (\mathbf{g}_{\mathrm{eff}}^{(t)})^{\mathsf{T}}\mathbf{M}_t^{-1}\mathbf{g}_{\mathrm{eff}}^{(t)} \ge \|\mathbf{g}_{\mathrm{eff}}^{(t)}\|_2^2 / \lambda_{\max}(\mathbf{M}_t) > 0$. Thus, the raw step is a strict ascent direction with respect to the sampled gradient estimate $-\mathbf{g}_{\mathrm{eff}}^{(t)}$ of $\Phi$ on the working set, where the estimate is formed from the residuals averaged over the two iterates after a backtracking event. Note that this property is preserved by the uniform trust-region scaling $\tau_t > 0$, whereas the subsequent projection of the inequality rows onto $\mathbb{R}_{\le 0}$ can clip an active row at zero, in which case the applied step is not guaranteed to be an ascent direction. Therefore, the ascent property is established for the raw direction, and it extends to the applied step for pure-equality problems, to which no projection is applied. In Case~2 ($k \in \mathcal{I}_{\mathrm{ineq}}\setminus\mathcal{A}_t$, where $\bar{C}_k^{(t)} \le 0$ and $\nu_k^{(t)} \ge -10^{-8}$), setting $[\mathbf{g}_{\mathrm{eff}}^{(t)}]_k = 0$ and decoupling the off-diagonal row/column $k$ of $\hat{\boldsymbol{\Sigma}}_C^{(t)}$ yields $\Delta\nu_{\mathrm{raw},k}^{(t)} = 0$, which preserves the complementary slackness $\nu_k^{(t+1)} = 0$ (within $10^{-8}$) under the cone projection onto $\mathbb{R}_{\le 0}$. Finally, writing $\boldsymbol{\Sigma}_{\mathrm{shrunk}}^{(t)} = \alpha\,\mathrm{diag}(\mathbf{S}_C^{(t)}) + \mathbf{V}\mathbf{V}^{\mathsf{T}}$ with $\mathbf{V} = \sqrt{(1-\alpha)/(S-1)}\tilde{\mathbf{C}}^{\mathsf{T}} \in \mathbb{R}^{K \times S}$ gives $\mathbf{M}_t = \mathbf{D}_t + \theta_t\mathbf{V}\mathbf{V}^{\mathsf{T}}$ for $\omega = 1$ when no inequality row is inactive, where $\theta_t = \kappa\hat{\beta}_{\mathrm{eff}}^{(t)}$ and $\mathbf{D}_t = \theta_t\alpha\,\mathrm{diag}(\mathbf{S}_C^{(t)}) + \lambda_t\mathbf{I}_K$ is a diagonal matrix with entries of at least $\lambda_t > 0$. Applying the Sherman--Morrison--Woodbury identity to $(\mathbf{D}_t + \theta_t\mathbf{V}\mathbf{V}^{\mathsf{T}})^{-1}$ reduces the matrix inversion to that of the $S \times S$ capacitance matrix $\mathbf{I}_S + \theta_t\mathbf{V}^{\mathsf{T}}\mathbf{D}_t^{-1}\mathbf{V}$, which requires $O(S^2 K + S^3)$ operations, i.e., $O(S^2 K)$ operations for $K \ge S$. Note that when inactive inequality rows are decoupled (Case~2), the same identity applies to the working block $[\mathbf{M}_t]_{\mathcal{W}\mathcal{W}}$ with the rows of $\mathbf{V}$ and the diagonal entries of $\mathbf{D}_t$ restricted to $\mathcal{W}$, and the inactive rows yield $\Delta\nu_{\mathrm{raw},k}^{(t)} = 0$ directly. In addition, under EMA smoothing ($\omega < 1$), $\hat{\boldsymbol{\Sigma}}_C^{(t)} = \sum_{j=0}^{t} \pi_j\boldsymbol{\Sigma}_{\mathrm{shrunk}}^{(j)}$ is a convex combination of the estimates of all iterations up to $t$. Therefore, the identity applies with $\mathrm{diag}(\mathbf{S}_C^{(t)})$ replaced by $\sum_j \pi_j\,\mathrm{diag}(\mathbf{S}_C^{(j)})$ and $\mathbf{V}$ replaced by the column-wise stack of the factors $\sqrt{\pi_j}\mathbf{V}_j$, and the capacitance matrix grows to $S(t+1) \times S(t+1)$. \qed

\subsection{Covariance shrinkage and EMA curvature memory}

Assume that $S$ spin configurations $\{\mathbf{s}^{(m)}\}_{m=1}^S$ are returned at outer iteration $t$, and that the constraint rows are normalized by $\sqrt{N}$ ($\mathbf{A} = \mathbf{A}_{\mathrm{raw}}/\sqrt{N}$, $\mathbf{B} = \mathbf{B}_{\mathrm{raw}}/\sqrt{N}$) such that the residual norms remain scale-consistent across $N$. Let $\mathbf{C}^{(m)} = \mathbf{A}\mathbf{s}^{(m)} - \mathbf{B}$ denote the $m$-th residual vector and $\bar{\mathbf{C}}^{(t)} = \frac{1}{S}\sum_{m=1}^S \mathbf{C}^{(m)}$ denote the ensemble mean. The centered residual matrix $\tilde{\mathbf{C}} \in \mathbb{R}^{S \times K}$ with rows $\tilde{\mathbf{c}}_m^{\mathsf{T}} = (\mathbf{C}^{(m)} - \bar{\mathbf{C}}^{(t)})^{\mathsf{T}}$ gives the empirical covariance $\mathbf{S}_C^{(t)} = \frac{1}{S-1}\tilde{\mathbf{C}}^{\mathsf{T}}\tilde{\mathbf{C}}$. To suppress the finite-sample Wishart eigenvalue dispersion when $S/K$ is modest, the off-diagonal entries of $\mathbf{S}_C^{(t)}$ are attenuated by means of linear shrinkage \cite{ledoit2004well} with a fixed weight $\alpha \in [0,1]$ toward the diagonal target $\mathrm{diag}(\mathbf{S}_C^{(t)})$, i.e.,
\begin{equation}
  \boldsymbol{\Sigma}_{\mathrm{shrunk}}^{(t)} = (1-\alpha)\,\mathbf{S}_C^{(t)} + \alpha\,\mathrm{diag}(\mathbf{S}_C^{(t)}).
  \label{eq:shrink}
\end{equation}
The shrinkage weight was set equal to $\alpha = 0.30$ in all of the \texttt{SelfCal-QSQN} runs with the exception of the $S=60$ Wishart stabilization benchmark (Fig.~\ref{fig:ablation}a), in which $\alpha = 0.35$ was used. Across consecutive iterations, the EMA smoothing is given by $\hat{\boldsymbol{\Sigma}}_C^{(t)} = (1-\omega)\hat{\boldsymbol{\Sigma}}_C^{(t-1)} + \omega\boldsymbol{\Sigma}_{\mathrm{shrunk}}^{(t)}$ with $\hat{\boldsymbol{\Sigma}}_C^{(0)} = \boldsymbol{\Sigma}_{\mathrm{shrunk}}^{(0)}$, where $\omega \in (0,1]$ denotes the memory weight. Note that the smoothing was activated only under low read budgets, namely with $\omega = 0.50$ in the $S=60$ Wishart stabilization benchmark and in the $S \le 250$ read-budget runs (Fig.~\ref{fig:ablation}a,b), including the Pegasus \texttt{Fixed-}$\beta$\texttt{ QSQN} curve in Fig.~\ref{fig:ablation}b. In all of the other runs, no memory was applied ($\omega = 1$), and thus $\hat{\boldsymbol{\Sigma}}_C^{(t)} = \boldsymbol{\Sigma}_{\mathrm{shrunk}}^{(t)}$.

\subsection{Secant self-calibration, KKT projection, and Woodbury solution}

Let the consecutive dual displacements be denoted by $\Delta\boldsymbol{\nu}^{(t-1)} = \boldsymbol{\nu}^{(t)} - \boldsymbol{\nu}^{(t-1)}$ and the observed residual shifts be denoted by $\mathbf{y}_t = \Delta\bar{\mathbf{C}}^{(t)} = \bar{\mathbf{C}}^{(t)} - \bar{\mathbf{C}}^{(t-1)}$. In accordance with equation~\eqref{eq:fdt}, a first-order Taylor expansion against the averaged shrunk covariance $\boldsymbol{\Sigma}_{\mathrm{avg}}^{(t)} = \frac{1}{2}(\boldsymbol{\Sigma}_{\mathrm{shrunk}}^{(t)} + \boldsymbol{\Sigma}_{\mathrm{shrunk}}^{(t-1)})$ gives $\mathbf{y}_t \approx +\hat{\beta}_{\mathrm{eff}}\mathbf{u}_t$, where $\hat{\beta}_{\mathrm{eff}}$ denotes the effective susceptibility gain and $\mathbf{u}_t = \boldsymbol{\Sigma}_{\mathrm{avg}}^{(t)}\Delta\boldsymbol{\nu}^{(t-1)}$. Projecting this relation onto the displacement $\Delta\boldsymbol{\nu}^{(t-1)}$ yields the secant estimate $\beta_{\mathrm{raw}}^{(t)} = (\Delta\boldsymbol{\nu}^{(t-1)\mathsf{T}}\mathbf{y}_t)/(\Delta\boldsymbol{\nu}^{(t-1)\mathsf{T}}\mathbf{u}_t)$. To prevent shot-noise corruption near stationarity, the secant updates are gated by the signal-to-noise ratio $\mathrm{SNR}_t = \|\mathbf{y}_t\|_2 / \sqrt{\mathrm{tr}(\mathbf{S}_C^{(t)})/S}$, where the denominator is the standard error of the ensemble mean (with $\mathrm{tr}(\mathbf{S}_C^{(t)})$ floored at $10^{-6}$). Specifically, if $\mathrm{SNR}_t > 2.5$, $\Delta\boldsymbol{\nu}^{(t-1)\mathsf{T}}\mathbf{y}_t > 0$, and $\Delta\boldsymbol{\nu}^{(t-1)\mathsf{T}}\mathbf{u}_t > 10^{-8}$, $\hat{\beta}_{\mathrm{eff}}^{(t)}$ is updated in accordance with $\hat{\beta}_{\mathrm{eff}}^{(t)} = (1-\gamma_\beta)\hat{\beta}_{\mathrm{eff}}^{(t-1)} + \gamma_\beta\,\mathrm{clip}(\beta_{\mathrm{raw}}^{(t)}, \beta_{\min}, \beta_{\max})$ with $\gamma_\beta = 0.40$, where $(\beta_{\min}, \beta_{\max}) = (2.5, 28.0)$ in Phase~2 and $(1.2, 30.0)$ in Phase~3. Otherwise, $\hat{\beta}_{\mathrm{eff}}^{(t)}$ is held constant. The estimator is initialized by a topology-level prior of $\hat{\beta}_{\mathrm{eff}}^{(0)} = 8.0$ on Zephyr and $4.2$ on Pegasus, which is divided by $\max(1, 0.85\,\ell_{\mathrm{c}})$ for the minor-embedded graphs with a mean chain length of $\ell_{\mathrm{c}}$. Note that the two priors were selected after Phase~1 and before Phases~2 and~3. The Pegasus prior follows the constant gain $\beta_{\mathrm{fixed}} \approx 4.25$ of \texttt{Fixed-}$\beta$\texttt{ QSQN}, which lies within the Phase-1 FDT probe range on Pegasus ($3.68\text{--}5.10$), whereas the Zephyr prior was set slightly above the Phase-1 FDT probe range on Zephyr ($6.23\text{--}7.65$) in order to avoid over-relaxation at the first step. Both priors were then held fixed across all Phase-2 and Phase-3 runs and the Phase-4 FDT probes. In addition, a non-monotone overshoot safeguard monitors the norm $r_t = \|\mathbf{r}_{\mathrm{KKT}}^{(t)}\|_2$ of the natural KKT residual, which is defined as $[\mathbf{r}_{\mathrm{KKT}}^{(t)}]_k = \bar{C}_k^{(t)}$ for $k \in \mathcal{I}_{\mathrm{eq}}$ and $[\mathbf{r}_{\mathrm{KKT}}^{(t)}]_k = \nu_k^{(t)} - \min(0, \nu_k^{(t)} - \bar{C}_k^{(t)})$ for $k \in \mathcal{I}_{\mathrm{ineq}}$, and thus reduces to $\|\bar{\mathbf{C}}^{(t)}\|_2$ for pure-equality problems. The safeguard is triggered whenever a shot-noise excursion satisfies $r_t > \max(1.20\,r_{t-1}, r_{t-1} + 0.40\sqrt{N/32})$ in Phase~2, or $r_t > \max(1.25\,r_{t-1}, r_{t-1} + 0.45\sqrt{N/32})$ in Phase~3. In this case, the controller backtracks halfway toward the preceding iterate ($\boldsymbol{\nu}^{(t)} \leftarrow \frac{1}{2}(\boldsymbol{\nu}^{(t)} + \boldsymbol{\nu}^{(t-1)})$, with $\mathbb{R}_{\le 0}^{K_{\mathrm{ineq}}}$ projection on the inequality rows) and averages the residual estimate accordingly ($\bar{\mathbf{C}}^{(t)} \leftarrow \frac{1}{2}(\bar{\mathbf{C}}^{(t)} + \bar{\mathbf{C}}^{(t-1)})$). Furthermore, $\hat{\beta}_{\mathrm{eff}}^{(t)}$ is multiplied by $1.40$ (capped at $30.0$) in Phase~2 and by $1.35$ (capped at $32.0$) in Phase~3 to damp the subsequent oscillations. Note that after a backtracking event, the reference quantities $(\boldsymbol{\nu}^{(t-1)}, \bar{\mathbf{C}}^{(t-1)}, \boldsymbol{\Sigma}_{\mathrm{shrunk}}^{(t-1)}, r_{t-1})$ of the secant estimate and of the overshoot test are retained from the most recent iteration without backtracking. Note also that the iterations are indexed from $t=0$ in Algorithm~1 and in the update formulas of Methods, whereas Results, the figures, the tables, and the checkpoint labels quoted in Methods count the iterations from $1$, i.e., iteration $t$ in Results corresponds to index $t-1$ in the update formulas.

\begin{table}[!htbp]
  \centering
  \captionsetup{labelformat=empty}
  \caption{\textbf{Algorithm 1 $\mid$ Closed-loop \texttt{SelfCal-QSQN} optimization protocol.}}
  \addtocounter{table}{-1}
  \footnotesize
  \begin{tabularx}{\linewidth}{@{}>{\raggedright\arraybackslash}X@{}}
    \toprule
    \textbf{Input:} Sparse Ising objective $(\mathbf{J}_0, \mathbf{c})$, linear constraints $(\mathbf{A}_{\mathrm{eq}},\mathbf{B}_{\mathrm{eq}}, \mathbf{A}_{\mathrm{ineq}},\mathbf{B}_{\mathrm{ineq}})$, reads $S$, iterations $T$, polish candidates $m_{\mathrm{cand}}$, residual-adaptive field trust bound $\Delta h_{\max}$, DAC rail $h_{\mathrm{rail}}$.\\
    \midrule
    1. \textbf{Static Mapping:} Map $\mathbf{J}_0$ once ($1$-to-$1$ native or short-chain minor-embedding) and initialize $\boldsymbol{\nu}^{(0)}=\mathbf{0}$ and the topology-level prior $\hat{\beta}_{\mathrm{eff}}^{(0)}$ ($8.0$ on Zephyr, $4.2$ on Pegasus).\\[2pt]
    2. \textbf{For} $t = 0, 1, \dots, T-1$ \textbf{do:}\\
    \hspace*{0.8em}a. Synthesize longitudinal fields $\mathbf{h}^{\mathrm{eff}}(\boldsymbol{\nu}^{(t)}) = \mathbf{c}-\mathbf{A}^{\mathsf{T}}\boldsymbol{\nu}^{(t)}$ (rescaled jointly with $\mathbf{J}_0$ by a common factor if any field exceeds $h_{\mathrm{rail}}$) and execute a single batch of $S$ QPU reads.\\
    \hspace*{0.8em}b. Evaluate $\bar{\mathbf{C}}^{(t)}$ and compute the shrunk covariance $\hat{\boldsymbol{\Sigma}}_C^{(t)}$ via equation~\eqref{eq:shrink} (with EMA smoothing under low read budgets).\\
    \hspace*{0.8em}c. If $t \ge 1$, $\mathrm{SNR}_t > 2.5$, $\Delta\boldsymbol{\nu}^{(t-1)\mathsf{T}}\mathbf{y}_t > 0$, and $\Delta\boldsymbol{\nu}^{(t-1)\mathsf{T}}\mathbf{u}_t > 10^{-8}$, update $\hat{\beta}_{\mathrm{eff}}^{(t)}$. If the residual exceeds the overshoot threshold, backtrack halfway to $\frac{1}{2}(\boldsymbol{\nu}^{(t)}+\boldsymbol{\nu}^{(t-1)})$ and boost $\hat{\beta}_{\mathrm{eff}}^{(t)}$.\\
    \hspace*{0.8em}d. Form active-set KKT gradient $\mathbf{g}_{\mathrm{eff}}^{(t)}$, adaptive damping $\lambda_t$ (from the residual norm measured prior to backtracking), and solve $(\kappa\hat{\beta}_{\mathrm{eff}}^{(t)}\hat{\boldsymbol{\Sigma}}_C^{(t)} + \lambda_t\mathbf{I}_K)\Delta\boldsymbol{\nu}_{\mathrm{raw}}^{(t)} = -\mathbf{g}_{\mathrm{eff}}^{(t)}$.\\
    \hspace*{0.8em}e. Apply local-field trust region $\tau_t = \min(1, \Delta h_{\max}/\|\mathbf{A}^{\mathsf{T}}\Delta\boldsymbol{\nu}_{\mathrm{raw}}^{(t)}\|_\infty)$ and project inequality rows onto $\mathbb{R}_{\le 0}^{K_{\mathrm{ineq}}}$.\\
    \hspace*{0.8em}f. Pass the $m_{\mathrm{cand}}$ lowest-violation raw samples through the deterministic 1-bit greedy polish ($O(NK)$ per flip) and update the best incumbent $\mathbf{s}^*$ without feedback to the dual update.\\[2pt]
    3. \textbf{Stage-2 Refinement:} Optionally run paused reverse annealing at locked $\boldsymbol{\nu}^*$, followed by the 1-bit greedy polish.\\
    \bottomrule
  \end{tabularx}
\end{table}

For the slack-free inequalities ($k \in \mathcal{I}_{\mathrm{ineq}}$, $\nu_k \le 0$), the active index set is defined as $\mathcal{A}_t = \{k \in \mathcal{I}_{\mathrm{ineq}} : \bar{C}_k^{(t)} > 0 \text{ or } \nu_k^{(t)} < -10^{-8}\}$. The active-set gradient is given by $[\mathbf{g}_{\mathrm{eff}}^{(t)}]_k = \bar{C}_k^{(t)}$ for $k \in \mathcal{I}_{\mathrm{eq}} \cup \mathcal{A}_t$. The inactive entries ($k \in \mathcal{I}_{\mathrm{ineq}}\setminus\mathcal{A}_t$) satisfy $[\mathbf{g}_{\mathrm{eff}}^{(t)}]_k = 0$, and the corresponding off-diagonal covariance couplings are set equal to zero. Based on the residual-adaptive Levenberg--Marquardt damping $\lambda_t = \lambda_0 + \gamma/(\tilde{r}_t + 0.5)$ with $\tilde{r}_t = r_t/\sqrt{N/32}$, $\lambda_0 = 2.0$, and $\gamma = 3.0$, which increases toward $8.0$ near feasibility to suppress finite-sample shot-noise jitter, the raw step is obtained by solving $(\kappa\hat{\beta}_{\mathrm{eff}}^{(t)}\hat{\boldsymbol{\Sigma}}_C^{(t)} + \lambda_t\mathbf{I}_K)\Delta\boldsymbol{\nu}_{\mathrm{raw}}^{(t)} = -\mathbf{g}_{\mathrm{eff}}^{(t)}$ with $\kappa = 1.10$. Note that $\kappa > 1$ provides a conservative margin against the over-relaxation that results from an underestimated gain. In all of the reported runs ($K \le 32$), the $K \times K$ system was solved directly. For $K \gg S$ and $\omega = 1$, the identical solution is given by the Woodbury formula
\begin{equation}
  \begin{aligned}
    (\mathbf{D}_t &+ \theta_t\mathbf{V}\mathbf{V}^{\mathsf{T}})^{-1}\mathbf{r} \\
    &= \mathbf{D}_t^{-1}\mathbf{r} - \theta_t\mathbf{D}_t^{-1}\mathbf{V}\bigl(\mathbf{I}_S + \theta_t\mathbf{V}^{\mathsf{T}}\mathbf{D}_t^{-1}\mathbf{V}\bigr)^{-1}\mathbf{V}^{\mathsf{T}}\mathbf{D}_t^{-1}\mathbf{r},
  \end{aligned}
  \label{eq:woodbury}
\end{equation}
where $\mathbf{D}_t = \theta_t\alpha\,\mathrm{diag}(\mathbf{S}_C^{(t)}) + \lambda_t\mathbf{I}_K$ is a positive diagonal matrix that collects the diagonal shrinkage target and the damping term, $\theta_t = \kappa\hat{\beta}_{\mathrm{eff}}^{(t)}$ weights the low-rank sample factor $\mathbf{V} = \sqrt{(1-\alpha)/(S-1)}\,\tilde{\mathbf{C}}^{\mathsf{T}} \in \mathbb{R}^{K \times S}$, and $\mathbf{r} = -\mathbf{g}_{\mathrm{eff}}^{(t)}$ denotes the right-hand side. The projected local-field shifts $\mathbf{A}^{\mathsf{T}}\Delta\boldsymbol{\nu}_{\mathrm{raw}}^{(t)}$ are bounded uniformly via $\Delta\boldsymbol{\nu}^{(t)} = \tau_t\Delta\boldsymbol{\nu}_{\mathrm{raw}}^{(t)}$ with $\tau_t = \min(1, \Delta h_{\max}/\|\mathbf{A}^{\mathsf{T}}\Delta\boldsymbol{\nu}_{\mathrm{raw}}^{(t)}\|_\infty)$, where the field trust bound is set at each iteration to $\Delta h_{\max} = \min(0.32, 0.06 + 0.05\,\tilde{r}_t)$ and thus tightens as the residual decays. Note that the residual norm $r_t$ that enters $\lambda_t$ and $\Delta h_{\max}$ is the value measured at iteration $t$ prior to any backtracking, whereas $\mathbf{g}_{\mathrm{eff}}^{(t)}$ and $\mathcal{A}_t$ are formed from the backtracked quantities. In addition, the bound applies to the step taken from the backtracked iterate, and thus the field change between two consecutive QPU programs can exceed $\Delta h_{\max}$ at a backtracked iteration. Note also that the uniform contraction preserves the exact multi-constraint cross-compensation direction in $\mathbb{R}^K$ prior to the dual-cone projection $\nu_k^{(t+1)} = \min(0, \nu_k^{(t)} + \Delta\nu_k^{(t)})$ ($k \in \mathcal{I}_{\mathrm{ineq}}$, Algorithm~1).

\subsection{Hardware execution, baseline calibration, and progressive stabilization settings}

All $891$ hardware experiments were executed via the D-Wave Ocean SAPI interface on \texttt{Advantage\_\allowbreak system6} (Pegasus P16, $5{,}612$ active qubits in the giant connected component, $40{,}088$ couplers, $h \in [-4,4]$) and \texttt{Advantage2\_\allowbreak system1} (Zephyr Z12, $4{,}560$ active qubits, $41{,}205$ couplers, $h \in [-4,4]$) \cite{dwave_zephyr2021,dwave_adv_perf,dwave_adv2_4400q,dwave_timing_doc}. Note that each run in Table~\ref{tab:campaigns} corresponds to one problem-instance record, and the QPU calls of all method executions stored in that record are counted in the QMI column. The $S$ reads of an iteration were acquired under one programmed Hamiltonian and were submitted as a single QMI when $S$ did not exceed the per-QMI read cap that the scripts derive conservatively from the $1.0\text{-s}$ run-duration limit of the solvers ($1{,}824$ reads at $t_a = 20\,\mu\mathrm{s}$ and $794$ reads at $t_a = 500\,\mu\mathrm{s}$). Otherwise, the reads were split into consecutive QMIs with the identical Hamiltonian, namely into two QMIs for $S=2{,}000$ and $S=3{,}000$, three QMIs for $S=4{,}000$ in the $\mathrm{SRT}=0$ FDT probes, two QMIs for $S=800$ at $t_a = 500\,\mu\mathrm{s}$, and two QMIs for each $400$-read reverse-annealing configuration. In the randomized gauge runs, each sub-batch was submitted as a separate QMI. The suites used the following reads per iteration $S$ and outer iterations $T$, namely $S=1{,}000$ and $T=22$ on Zephyr ($T=18$ on Pegasus) for the native scale-up, $S=2{,}000$ and $T=30$ at full chip, $S=800$ and $T=22$ for the high-$K$, annealing-schedule, non-native, and mixed KKT suites, $S=60$ and $T=25$ for the Wishart stabilization benchmark, $S=1{,}000$ and $T=24$ for the portfolio benchmark, and $T=20$ on Zephyr ($T=18$ on Pegasus) for the read-budget and SRT sweeps. The \texttt{Ohzeki} baselines of Phase~4 used the same $S$ and $T$ as the corresponding suites. In Table~\ref{tab:campaigns}, Phase~1 ($316$ runs) evaluated \texttt{Fixed-}$\beta$\texttt{ QSQN} prior to the introduction of online self-calibration or overshoot safeguarding. This phase comprises $90$ native scale-up runs (providing the Pegasus $N \le 4{,}096$ \texttt{Fixed-}$\beta$\texttt{ QSQN} column in Table~\ref{tab:scaling}), $64$ baseline calibration runs, $80$ preliminary \texttt{Fixed-}$\beta$ regularization sweeps, $2$ three-point $\mathrm{SRT}=2$ FDT probes ($N=128, K=4, S=2{,}500$, yielding $\beta_{\mathrm{emp}} = 3.68\text{--}5.10$ with $R^2$ up to $0.954$ on \texttt{Advantage\_\allowbreak system6} and $\beta_{\mathrm{emp}} = 6.23\text{--}7.65$ with $R^2$ up to $0.958$ on \texttt{Advantage2\_\allowbreak system1}), and $80$ exploratory \texttt{Fixed-}$\beta$ read-budget and schedule runs (including the Pegasus \texttt{Fixed-}$\beta$ curve in Fig.~\ref{fig:ablation}b). All \texttt{SelfCal-QSQN} runs were executed in Phases~2 and~3. Specifically, Phase~2 ($213$ runs) comprises $50$ high-$K$ runs (Fig.~\ref{fig:fdt}b) and $40$ annealing-schedule $t_a$ runs (Fig.~\ref{fig:fdt}c), each of which includes the paired \texttt{Fixed-}$\beta$\texttt{ QSQN} trajectory on the identical instance, together with $20$ progressive Wishart stabilization runs (Fig.~\ref{fig:ablation}a), $2$ six-point $\mathrm{SRT}=4$ FDT probes, $24$ full-chip runs (Fig.~\ref{fig:scaling}b and Table~\ref{tab:scaling}), $32$ two-stage portfolio runs (Fig.~\ref{fig:ablation}c), and $45$ matched-seed Zephyr $N \le 4{,}096$ \texttt{SelfCal-QSQN} scale-up runs (Table~\ref{tab:scaling}). Phase~3 ($94$ runs) comprises $40$ non-native sparse-graph runs (Fig.~\ref{fig:phase3}a), $30$ mixed KKT runs (Fig.~\ref{fig:phase3}b,c), and $24$ Zephyr $S$ and SRT runs (Fig.~\ref{fig:ablation}b). Finally, Phase~4 ($268$ runs) comprises the fixed-step and diminishing-step Ohzeki baselines across all six benchmark suites plus $2$ six-point $\mathrm{SRT}=0$ FDT probes (Fig.~\ref{fig:fdt}a). In the symmetrized finite-difference probes ($N=128, K=5, S=4{,}000$), $\beta_{\mathrm{emp}} = \mathrm{tr}(\mathbf{J}_{\mathrm{FD}})/\mathrm{tr}(\mathbf{S}_C)$ defines the slope of the zero-intercept covariance reference line plotted in Fig.~\ref{fig:fdt}a, and $R^2 = r^2$ denotes the squared Pearson correlation across the $K^2 = 25$ symmetrized matrix entries ($0.9916$ on Zephyr and $0.8755$ on Pegasus if restricted to the $15$ unique upper-triangular entries at $\mathrm{SRT}=0, t=3$). The zero-intercept fit $\mathbf{J}_{\mathrm{FD}} = \beta_{\mathrm{emp}}\mathbf{S}_C$ yields $R_0^2 = 1 - \|\mathbf{J}_{\mathrm{FD}}-\beta_{\mathrm{emp}}\mathbf{S}_C\|_F^2 / \|\mathbf{J}_{\mathrm{FD}}-\bar{J}\mathbf{1}\mathbf{1}^{\mathsf{T}}\|_F^2 = 0.992$ on Zephyr and $0.811$ on Pegasus under $\mathrm{SRT}=0$, and $0.958$ and $0.969$, respectively, under $\mathrm{SRT}=4$. In addition, $\beta_{\mathrm{dom}} = (\mathbf{v}_{\max}^{\mathsf{T}}\mathbf{J}_{\mathrm{FD}}\mathbf{v}_{\max})/(\mathbf{v}_{\max}^{\mathsf{T}}\boldsymbol{\Sigma}_{\mathrm{shrunk}}\mathbf{v}_{\max})$ defines the shrunk dominant-mode slope along the unit principal eigenvector $\mathbf{v}_{\max}$ of $\mathbf{S}_C$, where $\boldsymbol{\Sigma}_{\mathrm{shrunk}} = 0.70\mathbf{S}_C + 0.30\,\mathrm{diag}(\mathbf{S}_C)$. The $\mathrm{SRT}=4$ probes (Phase~2) used a finite-difference step of $\delta = 0.04\sqrt{N/32}$, whereas the $\mathrm{SRT}=0$ probes (Phase~4) used $\delta = 0.035\sqrt{N/32}$ on a separately extracted instance. Both probe trajectories followed the \texttt{SelfCal-QSQN} update over $18$ iterations without the overshoot safeguard, with the field trust bound capped at $0.32$ and $0.30$, respectively. Unless varied explicitly (Fig.~\ref{fig:fdt}c), the annealing duration was set to $t_a = 20\,\mu\mathrm{s}$ and \texttt{auto\_\allowbreak scale} was set to \texttt{False} for all of the dual schemes to preserve exact physical longitudinal calibration relative to $\mathbf{J}_0 \in [-1,1]$, whereas the embedded quadratic-penalty QUBO baseline was submitted with \texttt{auto\_\allowbreak scale=True}. In the event that a programmed value exceeded $99.5\%$ of the corresponding DAC range, $\mathbf{h}^{\mathrm{eff}}$ and $\mathbf{J}_0$ were rescaled jointly by a common factor. Note that, among the trajectories for which the scale factor was logged, this rescaling was triggered only in $38$ trajectories of the preliminary \texttt{Ohzeki} $\eta_t = 0.05$ step, and in none of the \texttt{SelfCal-QSQN}, \texttt{Fixed-}$\beta$\texttt{ QSQN}, or calibrated \texttt{Ohzeki} baseline runs. The scale factor was not logged for the $74$ minor-embedded trajectories of the Phase-1 calibration grid on the $3$-regular graphs, for the FDT probes, or for the Stage-2 reverse-annealing refinements. Gauge-locked runs ($\mathrm{SRT}=0$, default and Fig.~\ref{fig:fdt}a) submitted $(h_q, J_{uv})$ directly in the fixed identity gauge, whereas randomized gauge runs ($\mathrm{SRT}=2$ above and $\mathrm{SRT}=4$ in Phase-2 FDT and ``Impact of shot noise and reverse annealing'') partitioned reads into equal client-side sub-batches mapped via $(h_q, J_{uv}) \mapsto (\sigma_q h_q, \sigma_u\sigma_v J_{uv})$ under independent $\sigma_q \in \{-1,+1\}$, matching Ocean's \texttt{SpinReversalTransformComposite} \cite{dwave_solver_doc}. For the non-native $3$-regular/modular sparse graphs, a single static heuristic minor-embedding mapped each logical spin $i$ to a short ferromagnetic chain $\mathcal{C}_i$ with uniform longitudinal weights $w_{i,q} = 1/|\mathcal{C}_i|$ ($q \in \mathcal{C}_i$) and intra-chain coupling $-J_{\mathrm{chain}}$ determined by Ocean's \texttt{uniform\_\allowbreak torque\_\allowbreak compensation} utility (with a $1.5$ exception guard for isolated zero-coupler chains, which was never triggered in any hardware run). The static embeddings were obtained with \texttt{minorminer.find\_\allowbreak embedding} (a $60\text{-s}$ timeout and $15$ tries per call with at most three calls in Phases~3 and~4, and a $45\text{-s}$ timeout in the Phase-1 calibration grid), and broken chains were resolved by the default majority vote of Ocean's \texttt{unembed\_\allowbreak sampleset}. Note that the graph generators leave $2$ to $10$ vertices of degree $2$ in each of the $N=256$ and $N=384$ instances and in one of the $N=64$ seeds, and thus these graphs are $3$-regular up to these vertices.

Baseline hyperparameters were selected from the $64$-run Phase-1 calibration grid at $N=64, K=4$ across native subgraphs and random $3$-regular graphs on both processors ($16$ calibration seeds per group, generated with instance seeds that are disjoint from those of the $5$ benchmark seeds in Table~\ref{tab:scaling}). Note that the $N=64$ non-native benchmark instances in Table~\ref{tab:phase3} coincide with the $3$-regular calibration instances of seeds $0$--$4$. In addition, the $N=64$ native benchmark instances in Table~\ref{tab:scaling} share their native subgraphs with the native calibration instances of the same seed indices, whereas their $(\mathbf{J}_0, \mathbf{c}, \mathbf{A}, \mathbf{B})$ are drawn independently. For the quadratic-penalty QUBO baseline $f_0(\mathbf{s}) + \lambda_{\mathrm{pen}}\|\mathbf{A}\mathbf{s}-\mathbf{B}\|_2^2$ on $\sqrt{N}$-normalized constraints (embedded via the defect-aware clique embedder \texttt{minorminer.busclique} with torque-compensated $J_{\mathrm{chain}}$), $\lambda_{\mathrm{pen}} \in \{0.10, 0.25, 1.00\}$ was swept over $5$ matched calibration seeds. The clique embedding of $K_N$ was attempted for $N \le 256$ in the native suite and for $N \le 128$ in the non-native suite, and returned no embedding at $N=176$ and $256$ on Pegasus and at $N=128$, $176$, and $256$ on Zephyr. Note that the general heuristic embedder was not applied to the complete graphs. The larger sizes were skipped by a fixed size threshold in the scripts and were recorded as \texttt{EMBEDDING\_\allowbreak WALL\_\allowbreak EXCEEDED} without an attempt. Note also that for $N \ge 384$, $K_N$ requires at least $N(N-1)/2 \ge 73{,}536$ physical couplers, which exceeds the $40{,}088$ and $41{,}205$ couplers of the two processors. In addition, the penalty baseline has no outer iterations and was sampled in $4$ batches of $S$ reads ($3$ batches in the calibration sweep), compared with $T$ batches for the iterative dual schemes. On Pegasus, the sweep yielded native/$3$-regular ensemble residuals $\|\bar{\mathbf{C}}\|_2$ (and polished violations) of $4.70 / 2.80$ ($0.362 / 0.369$) at $\lambda_{\mathrm{pen}}=0.10$ ($J_{\mathrm{chain}} \approx 0.98\text{--}1.01$), $4.73 / 2.73$ ($0.333 / 0.353$) at $\lambda_{\mathrm{pen}}=0.25$ ($J_{\mathrm{chain}} \approx 2.24\text{--}2.35$), and $4.71 / 2.68$ ($0.362 / 0.353$) at $\lambda_{\mathrm{pen}}=1.00$ ($J_{\mathrm{chain}} \approx 8.82\text{--}9.25$). On Zephyr, the corresponding values were $8.20 / 6.25$ ($0.362 / 0.394$), $8.19 / 6.56$ ($0.403 / 0.557$), and $8.25 / 6.70$ ($0.579 / 0.742$), respectively. Note that the Zephyr residuals at $\lambda_{\mathrm{pen}}=0.25$ were $7.19 \pm 3.75$ (native) and $6.15 \pm 2.46$ ($3$-regular) across all $16$ Zephyr calibration seeds, whereas the $5$ benchmark seeds in Table~\ref{tab:scaling} gave $9.614 \pm 2.98$. The setting of $\lambda_{\mathrm{pen}} = 0.25$ minimized the Pegasus polished violations while avoiding the $J_{\mathrm{chain}} \approx 8.8\text{--}9.3$ DAC compression and Zephyr degradation observed at $\lambda_{\mathrm{pen}} = 1.00$, and $\lambda_{\mathrm{pen}}=0.10$ gave no ensemble gain. Accordingly, $\lambda_{\mathrm{pen}} = 0.25$ was fixed across all scales. The first-order Ohzeki baseline \cite{ohzeki2020breaking} ($\boldsymbol{\nu}^{(t+1)} = \boldsymbol{\nu}^{(t)} - \eta_t\bar{\mathbf{C}}^{(t)}$) was executed as standard open-loop subgradient updates without the closed-loop overshoot safeguard of \texttt{SelfCal-QSQN}, with $\mathbb{R}_{\le 0}$ projection on the inequality rows and the same per-iteration 1-bit greedy polish procedure with best-incumbent tracking. The symmetrized QPU response matrices ($K=4\text{--}5$) exhibit condition numbers $\kappa_{\mathrm{cond}} = \lambda_{\max}/\lambda_{\min} \approx 38\text{--}90$ and linear stability limits $\eta_{\mathrm{crit}} = 2/\lambda_{\max} \approx 0.014\text{--}0.036$. In the $N=64, K=4$ calibration sweep of the two tested fixed steps, $\eta_t = 0.05 > \eta_{\mathrm{crit}}$ diverged to $24.38\text{--}32.44$ (and to $269.68\text{--}284.77$ at full chip in preliminary Phase~1--3 sweeps), whereas $\eta_t = 0.01 < \eta_{\mathrm{crit}}$ avoided divergence ($1.12\text{--}1.28$). Accordingly, all benchmark tables and figures evaluate two primary first-order baselines with the read budgets and iteration limits of Phases~1--3, namely (i) the $K=4$-calibrated fixed step \texttt{Ohzeki} ($\eta_t = 0.01$) and (ii) the $K$-scaled $O(t^{-1/2})$ diminishing-step schedule \texttt{Ohzeki (Dim.)} with $\eta_t = 0.015 / [\max(1, K/4)\sqrt{1 + t/6}]$ for $t = 0, 1, \dots, T-1$. Note that the initial step of this schedule at $K=4$ ($0.015$) lies at the lower end of the $\eta_{\mathrm{crit}}$ range and decays below $0.014$ from $t=1$ onward. These baselines (Phase~4) were generated by the same instance generator with the same problem sizes and seeds as Phases~1--3. For each full-chip condition, the Phase 2 and Phase 4 runs were paired by solver, constraint count and instance seed. Both programs used the same seed formula and sorted the qubits and couplers in the same order. All 24 pairs agreed at the recorded precision in component size, coupler count, mean squared normalized constraint coefficient and planted-state energy. Given the archived Phase 2 Zephyr working graph, both programs reconstructed identical logical models for all 12 Zephyr cases and reproduced the statistics recorded in both phases. For the 12 Pegasus cases, only the archived summaries were available for comparison. The archive contains no Phase 4 working-graph snapshot for either solver, so graph identity across phases could not be verified independently. Within each Phase 2 full-chip job, \texttt{SelfCal-QSQN} and a separate fixed-step Ohzeki comparator ($\eta=0.05$) received the same generated model. For the non-native comparisons, the Phase 3 and Phase 4 programs regenerated identical logical models for all 40 solver--instance pairs. However, for the native suites below the full chip, namely the $N \le 4{,}096$ scale-up, high-$K$, portfolio, and mixed KKT suites and the $\mathrm{SRT}=0$ FDT probes, the Phase-4 subgraph extraction visited the neighbors of each qubit in sorted order and indexed the extracted qubits in sorted order, whereas Phases~1--3 visited the neighbors in seeded random order and retained the visiting order. As a result, the extracted native subgraphs, and thus the instances $(\mathbf{J}_0, \mathbf{c}, \mathbf{A}, \mathbf{B})$, differ between Phase~4 and Phases~1--3 in these suites. Therefore, the comparisons with the calibrated \texttt{Ohzeki} baselines in these suites are unpaired comparisons of multi-seed statistics over instances of the same generator, whereas the full-chip and non-native comparisons are paired. Note that the Phase-4 records store the instance metadata but not the instance matrices (see Data Availability).

In the progressive Wishart stabilization benchmark (Fig.~\ref{fig:ablation}a), all four pre-\texttt{SelfCal} variants solved $(\beta_{\mathrm{used}}\boldsymbol{\Sigma}_{\mathrm{shrunk}}^{(t)} + \lambda_t\mathbf{I}_K)\Delta\boldsymbol{\nu}_{\mathrm{raw}}^{(t)} = -\bar{\mathbf{C}}^{(t)}$ with $\beta_{\mathrm{used}} = \beta_{\mathrm{fixed}} = ((N/32)\sigma_A^2)^{-1} \approx 4.25$ ($\sigma_A^2$ denoting the mean squared entry of $\mathbf{A}$), diagonal shrinkage $\boldsymbol{\Sigma}_{\mathrm{shrunk}}^{(t)}$ in accordance with equation~\eqref{eq:shrink}, damping $\lambda_t = \lambda_0 + \gamma/(\|\bar{\mathbf{C}}^{(t)}\|_2/\sqrt{N/32} + 0.5)$, and coordinate-wise step clipping $|\Delta\nu_k^{(t)}| \le \Delta\nu_{\max}$, without EMA smoothing, secant calibration, or overshoot safeguarding. Note that the coordinate-wise clipping of these variants is not covered by the ascent statement of Proposition~\ref{prop:fdt}, which concerns the raw step and its uniform scaling. Specifically, \texttt{Unreg.\ FDT} inverted the unshrunk covariance ($\alpha = 0, \lambda_0 = 10^{-4}, \gamma = 0$) with $\Delta\nu_{\max} = 0.65\sqrt{N/32}$, whereas \texttt{Diag.\ Jacobi} ($\alpha = 1.0, \lambda_0 = 2.0, \gamma = 0$), \texttt{Fixed-Damp.} ($\alpha = 0.30, \lambda_0 = 2.0, \gamma = 0$), and \texttt{Fixed-}$\beta$\texttt{ QSQN} ($\alpha = 0.30, \lambda_0 = 2.0, \gamma = 3.0$) set $\Delta\nu_{\max} = \sqrt{N/32}\min(0.12, 0.03 + 0.02\|\bar{\mathbf{C}}^{(t)}\|_2/\sqrt{N/32})$. In the Stage-2 portfolio refinement, $\boldsymbol{\nu}^*$ was locked at the Stage-1 iterate with the lowest ensemble residual, and paused reverse annealing was seeded from each of the two best polished Stage-1 states at each of the two pause points $s_{\mathrm{pause}} \in \{0.48, 0.58\}$. Each of the resulting four configurations used $400$ reads under the schedule $[(0.0,1.0), (4.0\,\mu\mathrm{s}, s_{\mathrm{pause}}), (19.0\,\mu\mathrm{s}, s_{\mathrm{pause}}), (23.0\,\mu\mathrm{s}, 1.0)]$ with \texttt{reinitialize\_\allowbreak state=True} \cite{marshall2019power}, followed by the identical 1-bit greedy polish. The Stage-2 incumbent was initialized with the best polished Stage-1 state, and thus the reported Stage-2 result is the best incumbent over both stages. In the 1-bit greedy polish (labeled 1-flip in Fig.~\ref{fig:scaling} and Table~\ref{tab:scaling}), the $m_{\mathrm{cand}} = 12$ raw samples with the lowest constraint violation in each iteration are each refined by at most $\min(400, \max(20, \lfloor N/6 \rfloor))$ single-spin flips (with the cap of $400$ replaced by $350$ in Phases~3 and~4, and with $m_{\mathrm{cand}} = 10$ and at most $\min(256, \max(16, \lfloor N/8 \rfloor))$ flips in Phase~1). Each flip is chosen to yield the largest reduction in the constraint violation, with ties broken by the lowest objective change $\Delta f_0$ (evaluated over at most the first $24$ tied spins in Phases~2--4 and over all tied spins in Phase~1), and the refinement stops once the violation falls below $10^{-8}$ or no further reduction is possible. Note that the flip cap differs between \texttt{SelfCal-QSQN} ($400$) and the \texttt{Ohzeki} baselines of Phase~4 ($350$) only for $N \ge 2{,}106$, i.e., at $N=4{,}096$ and at full chip. Because the flip counts were not logged, the effect of this difference on the baseline feasibility rates at these sizes cannot be assessed. The best of the $m_{\mathrm{cand}}$ polished states is retained, and the best incumbent across all $T$ iterations is reported for every iterative dual scheme and for the symmetric classical baseline. Note that the polished states are not fed back to the dual update. Therefore, $\bar{\mathbf{C}}^{(t)}$ and $\hat{\boldsymbol{\Sigma}}_C^{(t)}$ are computed exclusively from the raw QPU reads. Classical Fortet--McCormick MILP benchmarks ($z_{ij} = s_i s_j \in [-1,1]$) were solved on the Zephyr seeds ($5$ seeds for $N \in \{32, 64, 128, 176, 256\}$ and the $6$ full-chip seeds at $N=4{,}560$) and $5$ Pegasus seeds ($N=32$) via \texttt{scipy.\allowbreak optimize.\allowbreak milp} (\texttt{HiGHS} \cite{huangfu2018highs,highs_web}, equality constraints imposed within a tolerance of $10^{-7}$, default relative optimality gap of $10^{-4}$, $5.0\text{-s}$ solver time limit). At full chip, no integer-feasible solution was returned within either the $5.0\text{-s}$ limit or an extended $14.0\text{-s}$ limit, which exceeds the $13.44\text{-s}$ estimated compute-only time of \texttt{SelfCal-QSQN}. Continuous Fortet--McCormick LP relaxation lower bounds $f_0^{\mathrm{LP}}$ were solved via \texttt{HiGHS}, namely with the default solver selection of \texttt{scipy.\allowbreak optimize.\allowbreak linprog} for $N \le 256$ and with interior-point optimization without simplex crossover at full chip. All bounds were evaluated on the archived chip graphs of the executed instances, including the $4{,}560$-qubit, $41{,}205$-coupler \texttt{Advantage2\_\allowbreak system1} working graph at full chip. For each of the six seeds, the reconstruction used the archived 4,560-node, 41,205-edge graph. The mean squared constraint coefficient and the planted-state energy agreed with the archived values within absolute tolerances of 0.000001 and 0.0001, respectively. In addition, the topology-invariant decoupled LP bound $\min_{\mathbf{A}_{\mathrm{raw}}\mathbf{s}=\mathbf{B}_{\mathrm{raw}}, \mathbf{s}\in[-1,1]^N}\mathbf{c}^{\mathsf{T}}\mathbf{s} - \sum_e |J_e|$ is reported as a looser bound that is independent of the edge assignment. In the symmetric classical 1-bit polish baseline (\texttt{HiGHS LP + 1-flip}), fractional spin marginals $x_i^{\mathrm{LP}} \in [0,1]$ from the \texttt{HiGHS} root LP relaxation were sampled across $T$ batches of $S$ Bernoulli draws (including deterministic threshold rounding) and passed through the identical $m_{\mathrm{cand}}=12$ vectorized 1-bit greedy polish using the fixed draw budget for $N \le 256$. At $N=4{,}560$, the two LP computations (decoupled and Fortet--McCormick) together took 5.06 $\pm$ 0.33 s on the reference host. Each recorded run then used a separate 1.5-s polish budget in batches of 2,000 draws, capped at 30 batches. The original records omit the number of completed batches. Therefore, the full-chip entry of this baseline depends on the host and is not matched to \texttt{SelfCal-QSQN} in the number of polish passes. All of the classical timings, namely the \texttt{HiGHS} runs and the outer-loop and 1-bit polish CPU times in Table~\ref{tab:scaling}, were measured on a single core of a cloud virtual machine (Intel Xeon Platinum 8481C processor with two logical cores, Python 3.11.2, NumPy 2.1.3, and SciPy 1.15.2, with the numerical libraries restricted to one thread). The outer-loop and polish CPU times were measured offline by replaying all $T$ outer-loop updates and all $T$ polish passes with a vectorized implementation on synthetic instances and sample batches of the matched size ($N$, $K$, $S$, and $T$), and were added to the archived pure QPU access times. The classical timings in Table~\ref{tab:scaling} and the time-limited MILP energies for $N \le 256$ are means over five repeated runs of the benchmark script on this host. The MILP incumbents were identical across the five runs with the exception of $N=64$, at which the $5$-seed mean energy varied between $-11.786$ and $-11.791$. At $N=256$, \texttt{HiGHS} was still processing the root node when the $5.0\text{-s}$ limit was reached on all of the $5$ seeds in each of the five runs. At this scale, the \texttt{HiGHS LP + 1-flip} baseline required $2.27\text{ s}$ on average over the $5$ seeds, of which the root LP solve accounted for approximately $96\%$. Note that the incumbents returned by \texttt{HiGHS} under a wall-clock time limit depend on the host and on the solver version. Therefore, the time-limited MILP energies and feasibility counts in Table~\ref{tab:scaling} are specific to this host. Across all multi-seed tables, figures, and text, $\pm\,\mathrm{s.d.}$ reports the population standard deviation ($\mathrm{ddof}=0$) over the $M$ independent seeds, matching the archived summary JSON records (multiplying by $\sqrt{M/(M-1)}$ yields the sample $\mathrm{s.d.}$ with denominator $M-1$).

\section*{Data Availability}
All $891$ per-run QPU experimental JSON records together with consolidated benchmark tables are archived in the experimental data packages. The records contain the per-iteration trajectories (the ensemble residual in all cases, together with the sample violation, polished violation, damping, calibrated gain, and step size where applicable), the pure QPU timing breakdowns, and all $44{,}066$ D-Wave SAPI \texttt{problem\_\allowbreak id} UUIDs. The problem Hamiltonians and constraint matrices are stored for the smaller instances of Phases~1--3 ($529$ of the $623$ records), the final dual multipliers are stored for Phases~3 and~4, and the empirical covariance and symmetrized finite-difference response matrices are stored for the FDT probes. Note that the Phase-4 records store the instance metadata only, and that the per-iteration multipliers and the native qubit assignments were not recorded. Furthermore, the chain-break fields in the archived records of the minor-embedded runs (\texttt{mean\_\allowbreak chain\_\allowbreak break\_\allowbreak fraction} and \texttt{chain\_\allowbreak break\_\allowbreak trajectory}) were not populated during the campaign (they hold the placeholder value of $0$), and thus they do not represent measured chain-break rates. In addition, the full-chip LP verification record, which archives the $4{,}560$-qubit, $41{,}205$-coupler \texttt{Advantage2\_\allowbreak system1} working graph of the full-chip instances, is included in the data and code packages, and thus the full-chip instances and the LP bounds can be regenerated offline. The full-chip campaign records contain per-iteration statistics. Individual spin configurations for each read were not retained. Prior to publication, these records are available from the corresponding author upon reasonable request.

\section*{Code Availability}
Prior to publication, the Python reference implementation, hardware campaign scripts, and classical verification programs are available from the corresponding author upon reasonable request. Upon publication, the code will be deposited in a public repository that assigns a permanent DOI, under a license restricted to academic and non-commercial use. The license conveys no rights under the patent application disclosed in the Competing Interests statement.

\section*{Acknowledgements}
This research received no specific external grant from funding agencies in the public, commercial, or not-for-profit sectors.

\section*{Author Contributions}
C.-H.O. conceived the proposed scheme, derived the fluctuation--dissipation dual curvature and online self-calibration identities, implemented the D-Wave Ocean SAPI execution software, conducted all 891 superconducting QPU experiments and classical MILP benchmarks, analyzed the data, and wrote the manuscript.

\section*{Competing Interests}
Chia-Ho Ou is the applicant and sole inventor named in U.S. Provisional Patent Application No.\ \texttt{\PatentAppNo}, filed with the United States Patent and Trademark Office, covering the self-calibrating susceptibility quasi-Newton constrained quantum annealing systems and methods reported in this study.

\end{document}